\documentclass[fleqn,usenatbib]{mnras}

\usepackage{newtxtext,newtxmath}

\usepackage[T1]{fontenc}
\usepackage{ae,aecompl}
\usepackage[utf8]{inputenc}
\usepackage{placeins}
\usepackage{float}
\usepackage{multirow}
\usepackage{verbatim}
\usepackage{adjustbox}

\usepackage{graphicx}	
\usepackage{tabularx}
\usepackage{amsmath}	
\usepackage{amsbsy}
\usepackage{natbib}
\usepackage{gensymb}
\usepackage{url}

\title[Environment of QSO triplets]{The environment of QSO triplets at $1 \lesssim z \lesssim 1.5$}

\author[M. C. Vicentin et al.]{
Marcelo C. Vicentin$^{1}$\thanks{E-mail: marcelo.vicentin@usp.br},
Pablo Araya-Araya$^{1}$,
Laerte Sodré Jr.$^{1}$,
Roderik Overzier$^{1,2}$, 
\newauthor \hspace{0.01cm} Eleazar R. Carrasco$^3$, Hector Cuevas$^4$
\\
$^{1}$Departamento de Astronomia, Instituto de Astronomia, Geof\'isica e Ci\^encias Atmosf\'ericas, Universidade de S\~ao Paulo \\ \hspace{0.01cm} Rua do Mat\~ao 1226 , Cidade Universit\'aria, 05508-900, S\~ao Paulo, SP , Brazil \\
$^{2}$Observat\'orio Nacional, Rua General Jos\'e Cristino, 77, S\~ao Crist\'ov\~ao, 20921-400, Rio de Janeiro, RJ, Brazil.\\ 
$^{3}$Gemini Observatory/NSF’s NOIRLab, Casilla 603, La Serena, Chile \\
$^{4}$Departamento de F\'isica y Astronom\'ia, Universidad de La Serena, Av. Juan Cisternas 1200 Norte, La Serena, Chile.}

\date{Accepted XXX. Received YYY; in original form ZZZ}

\pubyear{2020}
\hypersetup{draft}
\begin{document}
\label{firstpage}
\pagerange{\pageref{firstpage}--\pageref{lastpage}}
\maketitle

\begin{abstract}
We present an analysis of the environment of six QSO triplets at $1 \lesssim z \lesssim 1.5$ by analyzing multiband (\textit{r}, \textit{i}, \textit{z}, or \textit{g}, \textit{r}, \textit{i}) images obtained with Megacam at the CFHT telescope, aiming to investigate whether they are associated or not with galaxy protoclusters. This was done by using photometric redshifts trained using the high accuracy photometric redshifts of the \textit{COSMOS2015} catalogue. To improve the quality of our photometric redshift estimation, we included in our analysis near-infrared photometry (3.6 and 4.5$\mu$m) from the \textit{unWISE} survey available for our fields and the COSMOS survey. This approach allowed us to obtain good photometric redshifts with dispersion, as measured with the robust $\sigma_{NMAD}$ statistics (which scales as $(1 + z)^{-1}$), of $\sim$0.04 for our six fields. Our analysis setup was reproduced on lightcones constructed from the Millennium Simulation data and the latest version of the L-GALAXIES semi-analytic model to verify the protocluster detectability in such conditions. The density field in a redshift slab containing each triplet was then analyzed with a Gaussian kernel density estimator. We did not find any significant evidence of the triplets inhabiting dense structures,  such as a massive galaxy cluster or protocluster.
\end{abstract}




\section{Introduction}
\label{sec:intro}

In the hierarchical structure formation scenario of the current $\Lambda$CDM cosmological model,  small density fluctuations collapse to give rise to the first stars and galaxies. These galaxies continue to grow through the accretion or merger of other smaller or similar mass structures \citep[e.g., ][]{white78}, at the same time that the first groups and clusters of galaxies assemble \citep[e.g., ][]{kravtsov12}.

Studying the formation and evolution of galaxy clusters, by observations at different redshifts, is one of the main approaches to constrain our knowledge on the large-scale structures. At $z = 0$, cluster galaxies show the direct impact of living in such dense environments over billions of years. Examples include the high incidence of quiescent galaxies and the relationship between the morphological types and local galaxy density, indicating that the environment affects the evolution of a galaxy and its stellar population. Such characteristics are evidenced, for example, by the ``red sequence'' \citep{visvanathan77, bower92} in the colour-magnitude diagrams and the morphology-density relation \citep{dressler80}. Many studies seek to identify the molecular gas quenching level in different redshifts to infer the possible evolutionary stage of the galaxies. Those which are part of groups/clusters tend to lose their gas more quickly throughout their history and, consequently, no longer form stars. This can be caused, for example, by the interaction between different members of the group/cluster \citep[galaxy harassment, e.g., ][]{moore96}. Another possibility is the falling of a galaxy into a cluster, where the ICM causes a rapid gas depletion through ram pressure stripping \citep[e.g.,][] {gunn72}. There is also the phenomenon known as strangulation \citep[e.g.,][] {vandenbosch08}, where the stripping of the hot circumgalactic medium gas will interrupt the flow of cold gas into the galaxy, which ceases the fuel available for the formation of new stars. All of these processes are widely accepted as the main causes of star formation quenching and, consequently, reddening of galaxies that live in the densest environments in the universe \citep[galaxy quenching, e.g.,][]{kout19,liu19,trussler20, joshi20, gu20, lu20,tiley20}. 

At low redshifts ($z \lesssim $ 1), most high mass clusters are virialized structures that can be identified through a concentration of red galaxies\footnote{redMaPPer: \url{http://risa.stanford.edu/redmapper/}}. Additionally, their hot intra-cluster medium helps their identification through either their high X-ray emissions \citep{rosati02, mullis05, stanford06, mehrtens12} or by the Sunyaev-Zeldovich effect \citep{zeldovich72, bleem15, hilton18, huang20}.

At higher redshifts, $z \gtrsim 1$, most of the observed clusters are still in their formation process, and they are known as protoclusters \citep[see ][ for a review]{overzier16}. These structures harbour important clues about the galaxy formation process and the star formation history in such environments \citep{white91, bekki98, kodama98, vandokkum05, mei06}. However, their identification is not trivial. Since they do not present many of the observational properties of low-$z$, virialized clusters, the search for galaxy overdensities at high redshifts is one of the most effective ways to find them \citep{overzier16}. There are some galaxies or systems with special physical characteristics that may be used to find overdensities, such as Lyman Break Galaxies (LBGs) and Lyman Alpha Emitters (LAEs) \citep[e.g.,][]{overzier06, overzier08, chiang15, badescu17, higuchi19}, Hydrogen Alpha Emitters (HAEs) \citep{hatch11a, hayashi12}, submillimeter galaxies \citep{daddi09, capak11, rigby14, dannerbauer14}, radio galaxies \citep{venemans02,venemans07, hatch11b, hayashi12,  wylezalek13, cooke14}, isolated QSOs \citep{sanchez99,sanchez02,stott20} and systems of QSOs \citep{boris, onoue18}. These are all \textit{biased tracers} and are useful to find massive systems in the formation process \citep{overzier16}, since the bias between the distribution of baryonic matter and dark matter haloes in structure formation scenarios \citep{kaiser84} implies that high-z objects form in large high-z density fluctuations. 

Recently, \citet{stott20} studied 12 fields containing UV luminous QSOs at 1 < z < 2. The data were obtained with the \textit{Hubble Space Telescope} WFC3 G141 grism spectroscopy, and 2/3 of the sample showed significant galaxy overdensities associated with this type of quasar. \citet{cheng20} measured overdensities of 850$\mu m$-submillimetric sources in a sample of 46 protoclusters candidates selected from the Planck high-z catalogue \cite[PHz,][]{planck16} and the Planck Catalogue of Compact Sources (PCCS) that were followed up with Herschel-SPIRE \citep{planck15} and SCUBA-2 \citep{scuba17}, finding that 25 are associated with significant overdensities. 

To assess whether a given overdensity is real or a projected structure, it is necessary to analyze its redshift distribution, using either spectroscopic or photometric redshifts \citep[e.g.,][]{adami10, adami11, george14, wen11, jian14}. At $z >$ 1, spectroscopic redshift samples are quite limited or incomplete. However, the increasing number of photometric surveys in different broad, medium, and narrow-bands have made it possible to obtain high-quality photometric redshifts using template-fitting or machine learning techniques, useful for the search of high redshift galaxy clusters. For example, \citet{chiang14} reported 36 new candidates in the COSMOS field; by analyzing data from the Wide-field Infrared Survey Explorer (WISE) mission on the Pan-STARRS and SuperCOSMOS surveys, \citet{gonzalez19} found an amazing 1787 new high redshift clusters; at $z \sim 4$, \citet{toshikawa18} reported another 210 candidates in the wide layer area of the HSC-SSP; \citet{martinache18} also present new candidates for protoclusters by studying \textit{Herschel} and \textit{Planck} sources at 1.3 < z < 3.

It is still a matter of debate whether quasar systems are associated with overdense regions (here we will use  QSO or quasar indistinctly to designate objects with active nuclei). Previous studies with QSOs pairs have been inconclusive, with some finding association of these systems to protoclusters and others, not \citep[e.g., ][]{boris, farina11, green11, sandrinelli14, eftekharzadeh17, onoue18}. 

In this work, we analyze six fields containing QSOs triplets, where at least one of the members is a radio-loud object -- the radio signature at high redshift may be considered as an indication of a massive galaxy in a dense environment \citep[e.g.,][]{sanchez99, sanchez02, kuiper12, hatch}. The images were obtained with CFHT/Megacam in three different optical filters. We also include near-IR information from the \textit{unWISE} catalogue to improve the accuracy of photometric redshifts estimated through machine learning algorithms. 

The paper is organized as follows: in Section \ref{sec:obs_data} we present our observations and the reduction procedure, as well as the supplementary data (from the \textit{unWISE}, COSMOS, SDSS surveys, and simulated data) used in our analysis. In Section \ref{sec:analysis} we discuss our estimates of photometric redshifts, the analysis of the triplets environment through an evaluation of the galaxy overdensity significance field where they reside, and the protocluster detectability with the mock. These results are addressed in Section \ref{sec:discussion} and, finally, in Section \ref{sec:sum} we summarize our findings. Throughout this work we adopt a $\Lambda$CDM concordance cosmology, with recent cosmological parameters from \citet{planck1}: $h =$ 0.673, $\Omega_{\rm m}$ = 0.315 and $\Omega_{\rm \Lambda}$ = 0.685.


\section{OBSERVATIONS AND DATA REDUCTION}
\label{sec:obs_data}

In this section we describe how we have selected the sample of triplets for imaging at CFHT and the tools and procedures adopted for the image analysis, aiming to identify and extract the objects along with several photometric and morphological parameters. The detected objects were then corrected by galactic dust extinction and photometric calibrated in the SDSS system. We then discuss the star/galaxy separation, necessary for the analysis of the environment of the triplets, and finishes this Section describing the near-IR data extracted from the unWISE catalogue, which will be used later in this paper to improve estimates of photometric redshifts. 

\subsection{Sample selection}
\label{sec:sample_select}

In this work, we investigate six fields containing triplets of quasars. These systems were identified in the 13$^{th}$ edition of the \citet{veron10} quasar catalogue where we searched for quasar triplets in the redshift interval $1 \lesssim z \lesssim 1.5 $, with at least one object being radio-loud (RL), with separation between pairs in spectroscopic redshift less than 0.05, and in angular coordinates ($\theta$) less than 6 arcmin (between $\sim$ 6 and 8 cMpc for the redshift interval above, with a mean of 6.4 cMpc at the mean redshift of our six fields). Note that this angular separation criterion implies that the maximum distance between two QSOs in one triplet should be less than 12 arcmin. This constraint is 1 arcmin greater than the pair selection criteria of \citet{boris} and \citet{onoue18}, for quasars at z $\sim$ 1. Using a friends-of-friends algorithm, we found 21 systems, from which 6 were chosen for photometric follow-up taking advantage of an observational window at CFHT.

\begin{figure*}

	\includegraphics[width=\textwidth]{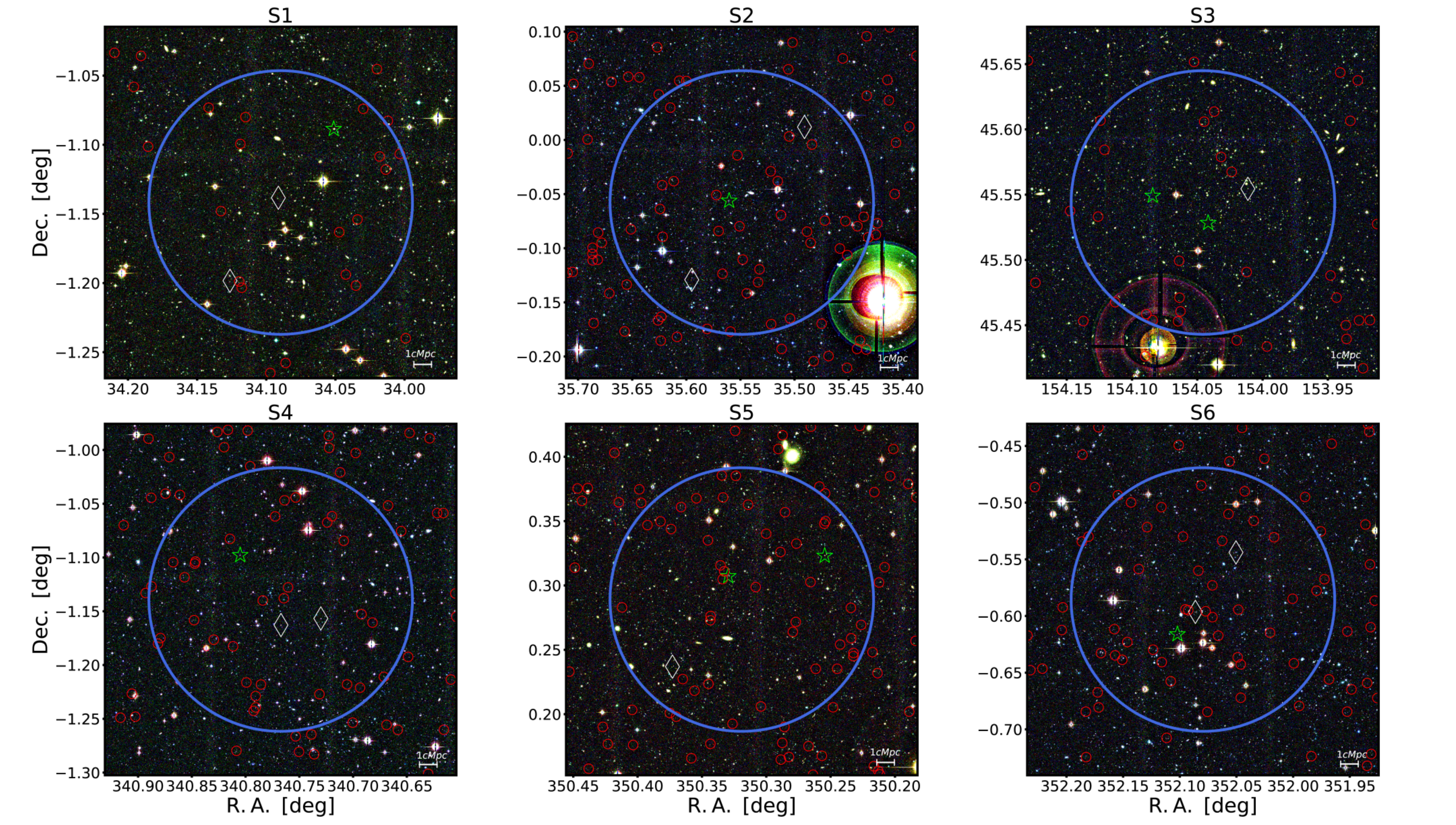}
    \caption{\small RGB composed images of our six fields with widths corresponding to 20$\times$20 cMpc. White diamonds and green stars stand for radio-quiet and radio-loud quasars, respectively; the blue circle has a projected radius of 7.5 cMpc;  red circles are galaxies with probability larger than 50\% to be in the redshift slab of the triplet (see section \ref{sec:density_field}).}

    \label{fig:rgb}
\end{figure*}

We obtained, for each triplet, multi-band images with MegaCam at the CFHT telescope\footnote{\url{https://www.cfht.hawaii.edu/Instruments/Imaging/Megacam/}} during the period 2014A (PI: Roderik Overzier). MegaCam is appropriate for this kind of work because it covers a large field-of-view, 1 x 1 square degree, with a resolution of 0.187 arcsecond per pixel. We used the medium dithering pattern, with a disk diameter of 30". In Figure \ref{fig:rgb}, we present a RGB composition of the central part of the images. Each field is centered on the triplet's centroid and has widths corresponding to 20$\times$20 cMpc.

In Table \ref{tab:obs_sample}, we present the seeing of the observations, the number and the exposure time for each imaging, and the corresponding total amount of time. Table \ref{tab:phys_sample} summarizes some additional relevant information: the field identification, the SDSS ID (J2000.0) of the triplet members, their redshifts, the difference between the mean and individual QSO redshifts ($\Delta v$), their reddening corrected magnitude in the reference/detection band, the mean redshift of the triplet, the magnitude interval in the reference magnitude used in our analysis (for the criteria, see Section \ref{sec:star_gal}), and the slab width in km/s (for the redshift slab definition, see Section \ref{sec:density_field}), the velocity intervals $\Delta v$ and $\Delta v_{\rm slab}$ are given as $c(\Delta z)/(1+\bar z)$ \citep{danese80}. We also included the minimum stellar mass ($M^{*}_{\rm min}$)\footnote{We consider as $M_{\rm min}^{*}$ the stellar mass of the percentile 1\% of the stellar mass distribution to avoid outliers.} that we are probing for each field within our magnitude limits and the redshift slabs of the quasars -- this information was obtained with the mock data (see Section \ref{sec:mocks_sample}). Finally, the bolometric luminosity and the virial black hole mass information, obtained by spectral analysis, were included from the \citet{shen11} catalogue\footnote{\url{http://quasar.astro.illinois.edu/BH\_mass/dr7.htm}} since massive and luminous quasars ($L_{\rm bol} \gtrsim 10^{15}$ erg/s and $M_{\rm BH} \gtrsim 10^{8}$ \(M_\odot\)) are more expected to be member of a greater structure (as we mentioned in Section \ref{sec:intro} citing other works). All quasars in our sample are above these values. Radio-quiet (RQ) objects in the \citet{veron10} catalogue are indicated in the table with `*'. In Appendix \ref{sec:apA}, we present the available QSOs spectra. For one object in S5 and two in S6, we do not have this information.

\begin{table*}
\centering
\caption{\small Sample observational information.}
\begin{adjustbox}{max width=\textwidth}
\begin{tabular}{lcccc}
\hline
Field               & Photometric band & Seeing (FWHM) (arcsec)   & n x Exposure time (s) & Total exposure time (s) \\ 
\hline
\multirow{3}{*}{S1} & r                & 0.66                   & 5 x 300               & 1500                 \\
                    & i                & 0.63                 & 6 x 310               & 1860                 \\
                    & z                & 0.62                    & 10 x 300               & 3000                 \\
\hline
\multirow{3}{*}{S2} & g                & 0.59                    & 6 x 300              & 1800                \\
                    & r                & 0.56                   & 6 x 300               & 1800                 \\
                    & i                & 0.57                   & 7 x 300               & 2100                 \\
\hline                    
\multirow{3}{*}{S3} & r                & 0.64                     & 5 x 300               & 1500                 \\
                    & i                & 0.64                    & 6 x 310               & 1861                 \\
                    & z                & 0.64                     & 3 x 300               & 900                  \\
\hline
\multirow{3}{*}{S4} & g                & 0.66                    & 6 x 300               & 1800                 \\
                    & r                & 0.51                      & 6 x 300           & 1800                     \\
                    & i                & 0.69                    & 7 x 300               & 2100                 \\
\hline
\multirow{3}{*}{S5} & r                & 0.49                   & 5 x 300               & 1500                 \\
                    & i                & 0.46                    & 12 x 310               & 3720                 \\
                    & z                & 0.53                    & 10 x 300               & 3000                 \\
\hline
\multirow{3}{*}{S6} & g                & 0.57                    & 6 x 300               & 1800                 \\
                    & r                & 0.51                     & 6 x 300               & 1800                 \\
                    & i                & 0.55                    & 7 x 300               & 2100   \\
\hline
\end{tabular}%
\end{adjustbox}
\label{tab:obs_sample}
\end{table*}

\begin{table*}
\centering
\caption{\small Physical information of the triplet sample. From the left to the right, the columns are: the field identification, the SDSS ID (J2000.0) of the triplet members, their redshifts, the mean redshift of the triplet, the difference between the mean and individual QSO redshifts, their reddening corrected magnitude
in the reference/detection band, the magnitude interval in the reference magnitude used in our analysis, the minimum stellar mass that we are probing for each field within our magnitude limits, the width of the redshift slabs of the quasars, the bolometric luminosity, and the virial black
hole mass. More details are presented in Section \ref{sec:sample_select}.} 
\begin{adjustbox}{max width=\textwidth}
\begin{tabular}{ccccccccccc}
\hline
Field & QSO & $z_{\rm QSO}$ &  $\bar z$ & $\Delta v [km/s]$ & Reference Band Magnitude & Adopted Limits & $log(M_{\rm min}^{*})$ [\(M_\odot\)] & $\Delta v_{\rm slab}$ [km/s]& $log(L_{\rm bol})$ (erg/s) & $log(M_{\rm BH})$ [\(M_\odot\)]
\\

\hline

\multirow{3}{*}{S1} & SDSSJ021612-010519  & 1.480 & \multirow{3}{*}{$1.506$} &  -3112  & $i = 17.325$  & \multirow{3}{*}{$17 < i < 23.5$}  &   \multirow{3}{*}{9.589}  &   \multirow{3}{*}{$\pm 15000$} & 47.04 & 9.22   \\
                    & SDSSJ021622-010818* & 1.518 &   & +1437 & $i = 19.956$             &                               &                            &     & 46.01 & 9.14     \\
                    & SDSSJ021630-011155* & 1.521 &  &  +1796  & $i = 20.205$             &                                 &                            &    & 45.96 &  8.98    \\
\hline

\multirow{3}{*}{S2} & SDSSJ022158+000043*    & 1.042 & \multirow{3}{*}{$1.053$} & -1607   & $g = 18.790$   & \multirow{3}{*}{$18.5 < g < 24.5$} & \multirow{3}{*}{8.995} & \multirow{3}{*}{$\pm$ 12582} & 46.29 & 9.54  \\
                    & SDSSJ022214 -000322   & 1.066 &  & +1900  & $g = 19.250$             &                                &                           &    & 46.04 & 9.33     \\
                    & SDSSJ022223-000745*    & 1.051  &  &   -292  & $g = 21.095$             &                              &                            &     & 45.65 & 8.87       \\

\hline

\multirow{3}{*}{S3} & SDSSJ101603+453316*    & 1.369 &  \multirow{3}{*}{$1.376$} & -884   & $r = 19.923$  & \multirow{3}{*}{$18 < r < 24$} & \multirow{3}{*}{9.267} & \multirow{3}{*}{$\pm$ 15326}  & 45.95 & 8.92  \\
                    & SDSSJ101610+453142    & 1.380  &  & +505  & $r = 19.039$             &                              &                     &        & 46.36 & 9.24             \\
                    & SDSSJ101620+453257    & 1.379  &  & +379   & $r = 17.626$             &                               &                     &    & 46.93 & 9.53
                                    \\

\hline

\multirow{3}{*}{S4} & SDSSJ224255-010924* & 1.033 & \multirow{3}{*}{$1.044$} & -1615  & $g = 19.347$  & \multirow{3}{*}{$18 < g < 24.5$} & \multirow{3}{*}{9.197} & \multirow{3}{*}{$\pm$ 10779} & 45.83 & 8.82 \\
                    & SDSSJ224304-010946* & 1.044 &   & 0   & $g = 19.269$  &                                   &                           &  & 45.96 & 8.99 \\
                    & SDSSJ224313-010552    & 1.054 &  &  +1615  & $g = 19.342$                                &                 &                  &    & 46.12 &9.21            \\
                    
\hline

\multirow{3}{*}{S5} & SDSSJ232101+001923   & 1.362 & \multirow{3}{*}{$1.352$} & +1275   & $i = 18.881$  & \multirow{3}{*}{$17.5 < i < 24$} & \multirow{3}{*}{9.315}  & \multirow{3}{*}{$\pm$ 16715}   & 46.34 & 8.70 \\
                    & SDSSJ232119+001826   & 1.327  &  & -3189 & $i = 20.358$                          &                                &               &    & - & -             \\
                    & SDSSJ232129+001413*     & 1.368 &  & +2041   & $i = 20.320$                         &                      &               &       & 45.58 & 8.80                  \\
                    
\hline

\multirow{3}{*}{S6} & SDSSJ232812-003238*    & 1.083 &  \multirow{3}{*}{$1.116$} & -4679   & $g = 20.997$  & \multirow{3}{*}{$18 < g < 24.5$} & \multirow{3}{*}{9.473}  & \multirow{3}{*}{$\pm$ 10219} & - & - \\
& SDSSJ232821-003547*  & 1.136 &  & +2835    & $g = 20.280$  &   &   &  & - & -  \\
& SDSSJ232824.5-003658 & 1.128 &  & +1701    & $g = 20.116$  &   &   &   & 45.79 & 8.35  \\
\hline

\end{tabular}
\end{adjustbox}

       {\tiny ~ \\
          \flushleft{Note: Radio-quiet QSOs are denoted by "*".} \\
         }
\label{tab:phys_sample}
\end{table*}


\subsection{Object extraction}

The CHFT science frames were processed with the package THELI \citep{erben05,schirmer13}. The images were bias/overscan-subtracted, trimmed, flat-fielded, and registered to a common pixel and sky coordinate positions using SCAMP \citep{bertin06}. A combined astrometric solution for all three filters (g, r, i, or r, i, z) was derived using the SDSS DR9 catalog. The resulting astrometric calibrated individual frames were then sky subtracted, re-sampled to a common position, and stacked with SWARP \citep[][ SWarp: Resampling and Co-adding FITS Images Together, Astrophysics Source Code Library]{bertin10}.

Then, we ran \texttt{SExtractor} \citep{bertin96} along with \texttt{PSFex} \citep{bertin2}, which performs PSF fitting photometry, on the combined frames of each band, to detect and extract objects and then create the photometric catalogues containing all the measurements that are necessary for this work: celestial coordinates (RA and DEC, J2000), photometric and morphological information such as magnitudes and their errors, the image width \texttt{FWHM}, \texttt{CLASS\_STAR}, the maximum surface brightness \textit{$\mu_{\rm max}$} and the \texttt{SPREAD\_MODEL}.
The \texttt{CLASS\_STAR} parameter is a point/extended source classifier that estimates the a posteriori probability that a detection made by \texttt{SExtractor} is a point source. \texttt{CLASS\_STAR} relies on a multilayer feed-forward neural network trained using supervised learning to generate the estimates. Objects with \texttt{CLASS\_STAR} close to 1 are likely point-sources, whereas those close to 0 are probably galaxies.

\texttt{SPREAD\_MODEL} is a powerful model-based morphological parameter obtained with \texttt{PSFex}. It gives a local normalized point spread function (PSF) for each detection, and indicates if it is best matched by a point-source model (\texttt{SPREAD\_MODEL} = 0), or by a “fuzzier” model, describing an extended object (\texttt{SPREAD\_MODEL} > 0) \texttt{PSFex} does not work directly on images. Instead, it operates on \texttt{SExtractor} catalogues, which have a small sub-image (``vignette”) recorded for each detection. This makes things much easier since one does not have to handle the detection and deblending processes. The catalogue files read by \texttt{PSFEx} must be in the \texttt{SExtractor} \texttt{FITS\_LDAC} binary format, which allows the software to have access to the original image header content. In order to use this tool, we performed the following steps for each field/band:

\begin{enumerate}
    \item Run \texttt{SExtractor} in \textit{single mode} with only the necessary parameters that will be used by \texttt{PSFex}. The result is a catalogue (\texttt{FITS\_LDAC} binary format file) containing all the identified objects with the parameters information.
    
    \item  Run \texttt{PSFex} to generate the \texttt{PSF} for each object. In this step, the output of (i) is used as  input, and the result will be the ``\texttt{.psf}'' files.
    
    \item Run \texttt{SExtractor} again including the output of (ii), which is responsible for the \texttt{SPREAD\_MODEL} parameter and its corresponding error. Also, once we know which filter has detected more objects (from (i)), we set \texttt{SExtractor} in \textit{dual mode} to use this filter as reference for the others. 
\end{enumerate}

\subsection{Complementary databases}

Besides the CFHT images, we have used some complementary databases to do the photometric calibration and to perform estimates of photometric redshifts (photo-zs). They are described below.


\subsubsection{SDSS}

The photometric calibration was made using the \textit{ugriz SDSS} system \citep{fukugita96} since our images are in similar filters. This survey has a photometric accuracy of 2-3\% and astrometric accuracy better than 0.1" \citep{pier03}.

We have used a query search in the \textit{Catalog Archive Server Jobs System (CasJobs)}\footnote{\url{https://skyserver.sdss.org/casjobs/}} for \textit{SDSS-DR14} to find the unsaturated stars in the same area of our images. The catalogues generated by \textit{CasJobs} contain the coordinates of each object and the photometric measurements in the same filters of our images. We present in Section \ref{sec:calibration} details of the photometric calibration.


\subsubsection{COSMOS2015}
\label{sec:cosmos2015}

In Section \ref{sec:photo-zs} we present our machine-learning approach to estimate photometric redshifts for our 6 fields. To train our algorithm, we have used data from the  \textit{COSMOS2015} \citep{laigle16} catalogue, supplemented with unWISE photometry.
\textit{COSMOS2015}  is a public catalogue with more than a million objects over a 2deg$^{2}$ of the \textit{COSMOS} field. The main motivation to use photometric redshifts from this catalogue to train our own photo-z estimator is because half of its objects have very accurate photo-zs, $\sigma_{\rm{\Delta z}}/(1+z_{\rm s}) \simeq 0.007$, through a comparison with the \textit{zCOSMOS}-bright spectroscopic redshifts (spec-zs). This high accuracy is due to a large number of photometric bands (>30) obtained by several surveys in this area, from the UV to the near-IR (e.g., UltraVISTA-DR2, Subaru/Hyper-Suprime-Cam, IRAC/\textit{Spitzer} (SPLASH)). 

This catalogue provides magnitudes in filters (B, V, r, i+, and z++, from Suprime-Cam/Subaru) similar to those of our images, with a depth of $i \sim 26.5$. In the case of the g filter, we applied the transformation proposed by \citet{jester05}, using apparent magnitudes in the B and V bands:
\begin{equation}
    g = V + 0.64 \times (B - V) - 0.13
    \label{eq_g_transf}
\end{equation}

We also (re)calibrated the magnitudes of \textit{COSMOS2015} in bands equal or similar to those of our fields in the \textit{SDSS} photometric system, as described in Section \ref{sec:calibration}. 

\subsubsection{Simulated Data}
\label{sec:mocks_sample}

The main goal of this work is to explore the QSO triplets environment. To quantify the possibility to detect protoclusters at their redshifts, given our observational constraints, we construct protocluster-lightcones (hereafter, \textit{PCcones}), using the technique presented in \citet{araya20}. These particular mocks consist of structures from the Millennium Simulation \citep{springel05} placed in the center of the line-of-sight of the mocks at a certain redshift. The \texttt{L-GALAXIES} \citep{henriques15} semi-analytical model is used to populate the dark matter halos of the Millennium simulation. The original area of each lightcone is $\pi$ deg$^2$, but we constrained them to 1 deg$^2$, which is the field-of-view of our observations. 

We placed $20$ different structures at the center of the lightcone in $4$ different redshifts: $z= 1.04$, $1.12$, $1.36$ and $1.51$. Following the definition of \citet{chiang13} for protocluster's type based on the mass that the cluster will have at z = 0, we include 8 Fornax-type ($M_{z=0} = 1.37-3.00 \times 10^{14} \ M_{\odot}$ ), 6 Virgo-type ($M_{z=0} = 3-10 \times 10^{14} \ M_{\odot}$) and 6 Coma-type ($M_{z=0} \geq  10^{15} \ M_{\odot}$) protoclusters at these redshifts.

Our observational data is not homogeneous in the sense that we have different photometric bands for each field. Additionally, the magnitude limits are not the same for the whole sample (see Table \ref{tab:phys_sample} for each field limits, and Section \ref{sec:star_gal} for the criteria). To address these differences, we have mimic real observations from the photometric catalogs of our $80$ lightcones by applying to them the adopted magnitude limits presented in Table \ref{tab:phys_sample}. Some fields contain the QSO triplet at similar redshifts. For example, the average redshifts of the QSO triplet in the S2 and S4 fields are at z $\sim$ 1.04. This also happens for S3 and S5, where the QSO systems are at z $\sim$ 1.36. Then, from each PCcone with placed structures at z = 1.04, we generated two mock photometric catalogs, which emulate both the S2 and S4 observations. We made the same (obtaining two mock catalogs from one PCcone) for PCcones with placed structures at z = 1.36 to mimic the S3 and S5 fields. In summary, we have 20 lightcones for each field (a total of 120 mocks after constraining the PCcones by the magnitude limits). The minimum stellar mass (defined as the 1\% percentile of the simulated mass distribution) probed for each field and redshift after taking into account the same constraints of the observations ranges from $log(M^{*}_{min}/M_{\odot}$) $\sim$ 9 to 9.6 (See Table \ref{tab:phys_sample}).


\subsection{Correction of the galactic extinction}

The magnitudes generated by \texttt{SExtractor} were first corrected by the extinction caused by the Milky Way's dust, which causes a slight increase in the magnitude of each object, depending on its celestial coordinates. 
This correction was determined using a routine in python, which reads the value of E(B-V) corresponding to the coordinates of each object on the \citet{schlegel98} maps, and calculates the extinction with the \citet{cardelli89} law for each photometric band\footnote{The packages used were \texttt{sfdmap} and \texttt{extinction}}. These extinctions were then subtracted from the corresponding magnitude for each object. 

\subsection{Photometric calibration}
\label{sec:calibration}

The photometric calibration was done using isolated and non-saturated stars in the \textit{SDSS} Data Release 14. We have ran a \texttt{SQL} script at \textit{CasJobs} using appropriate flags. After, we applied additional constraints to this data, as illustrated in Table \ref{tab:cal}, to ensure that the selected stars have indeed good photometry. 

\begin{table}
	\centering
	\caption{\small Constraints applied for selecting unsaturated stars for the photometric calibration.}
	\begin{tabular}{cc}
      \hline
      Parameter & \textit{Selection} \\
      \hline
     $\mu_{\rm max}$  &  $\mu_{\rm sat} + 0.5 < \mu_{\rm max} < \mu_{\rm sat} + 3.5$ \\
     \texttt{FWHM}  &  $<$ 4.2 \\
     \texttt{CLASS\_STAR}  &  $> 0.95$ \\
     \texttt{SPREAD\_MODEL}  &  $< 0.0002$ \\
     \texttt{MAG\_ERR}  & $<$ 0.03 \\
   	\hline
	\end{tabular}
	\label{tab:cal}
\end{table}

\begin{figure*}

	\includegraphics[width=\textwidth]{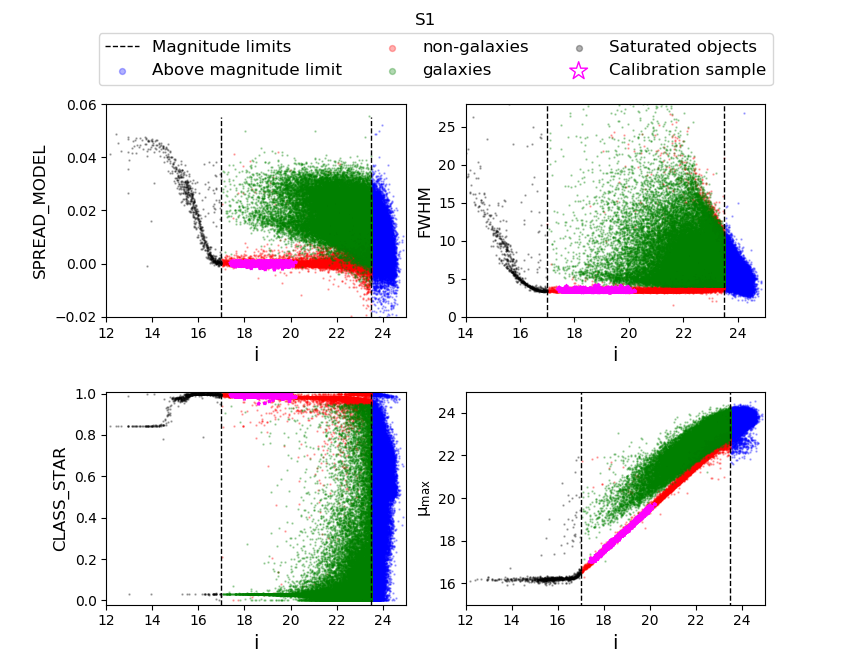}
    \caption{\small Object selection for the S1 field. Green objects are considered galaxies; red are non-galaxies; pink are the stars selected for calibration; black are objects considered as saturated, and; blue objects are those above the upper magnitude limit (see text).}

    \label{fig:star_gal}
\end{figure*}

These criteria were based on visual inspection of diagrams that relate morphological/photometric parameters to the magnitudes of the objects (see Figure \ref{fig:star_gal} for an illustration of the procedure applied to one of our fields). To eliminate the saturated objects we chose a lower magnitude limit using the $\mu_{\rm max}$ parameter (\textit{bottom-right} in this figure) since the saturated objects at the brighter magnitude-end have essentially a constant value for this parameter ($\mu_{\rm sat}$). This value defines the lower and upper values for $\mu_{\rm max}$ according to Table \ref{tab:cal}. We also set limits for other morphological parameters to make sure that would be no inconsistencies in the selection. 

The next step was a cross-match of celestial coordinates between our fields and  \textit{SDSS}, assuming a tolerance in the angular separation of the sources $\leq 1$ arcsec, and a 3$\sigma$ clipping to remove the outliers. Our final sample for calibration can be checked in the diagrams of Figure \ref{fig:star_gal} (pink objects).

\begin{figure}

	\includegraphics[width=\columnwidth]{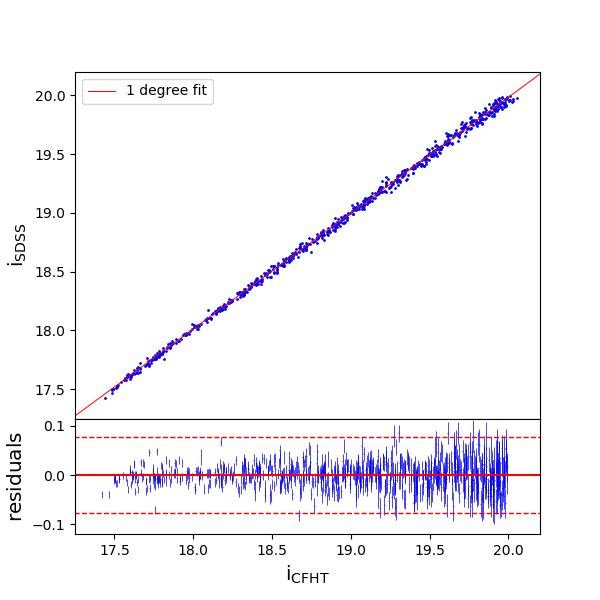}
    \caption{\small Photometric calibration for the i-band in the S1 field. Above: the linear calibration applied in this case. Below: the residuals of the fit as a function of the calibrated magnitude. The dashed horizontal red lines represent $\pm 3 <\sigma>$.}

    \label{fig:cal}
\end{figure}

After these steps, we transformed our instrumental magnitudes to the \textit{SDSS} photometric system using a linear fit. Figure \ref{fig:cal} shows this calibration for one of our fields. These transformations were applied to all bands in all fields, including the corresponding magnitudes of the \textit{COSMOS2015} catalogue. Typical errors in this calibration are in the range 0.02 to 0.04 for r, i, and z bands; and, 0.04 to 0.08 for the g band.

\subsection{Star/galaxy separation}
\label{sec:star_gal}

Given the objectives of this work, we need to identify the galaxies in our fields, and, for this, any point source present in our catalogues will be considered as a ``non-galaxy''. In this section, we describe our approach to galaxy selection.

\begin{figure}

	\includegraphics[width=\columnwidth]{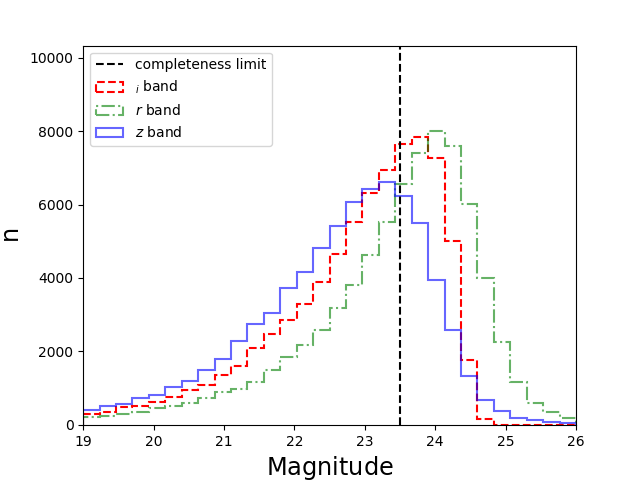}
    \caption{\small Magnitude distributions for the S1 field. Objects with reference band magnitude (in this case, the \textit{i} band) fainter than the magnitude completeness limit (the black dashed line; for criterion, see text) are discarded.} 

    \label{fig:upper_limit}
\end{figure}

First of all, we defined the magnitude interval of interest, considering limits due to image saturation at the bright side ($mag_{\rm{ref, min}}$, such that $\mu_{\rm max} = \mu_{\rm sat} + 0.5$ in the reference band) and the upper limit at the faint side as $\sim$0.1 below the peak of the reference band magnitude distribution; see Figure \ref{fig:upper_limit}. These limits are also presented in Table \ref{tab:phys_sample} for the reference band of each field.

Within this magnitude interval in the reference band, we consider galaxies as extended objects, identified using three morphological parameters, \texttt{SPREAD\_MODEL}, \texttt{CLASS\_STAR}, and \texttt{FWHM} (pixel), as a function of the magnitude in the reference band.

As mentioned before, the \texttt{SPREAD\_MODEL} parameter provides a powerful measurement of the width of an object, so we used it along with its error. By taking into account this error, the efficiency of the galaxy selection (true galaxies classified as such, over total true galaxies) increases considerably at the faint-end. This efficiency was measured in a controlled sample -- where the information whether the object is a star or a galaxy is known \footnote{\url{https://cdcvs.fnal.gov/redmine/projects/des-sci-verification/wiki/A\_Modest\_Proposal\_for\_Preliminary\_StarGalaxy\_Separation}}. 

We reinforce this selection by adopting limits for the \texttt{CLASS\_STAR} and \texttt{FWHM} parameters based on the \textit{star sequence} identified in plots like those shown in Figure \ref{fig:star_gal}, where the sequence of unsaturated stars can be well distinguished through visual inspection at low values of these parameters. Since \texttt{CLASS\_STAR} estimates the probability that an object is a point source, we expect that extended objects avoid high values for this parameter. Besides, the magnitude-$\mu_{\rm max}$ diagram also presents nice discrimination between extended and point-source objects. Based on the distribution of the objects in Figure \ref{fig:star_gal}, we have set the following limits on these parameters for an object to be considered as a galaxy: 
\\

$(\texttt{SPREAD\_MODEL} + 3 \times \texttt{SPREADERR\_MODEL}) \geq 0.003$  

\textbf{and} $\texttt{CLASS\_STAR} \leq 0.95$ 

\textbf{and} $\texttt{FWHM} \geq 4$ 
\\

The different colours in Figure \ref{fig:star_gal} show the different classes of objects after applying these constraints.
We also show in this figure the resulting classification in the magnitude-$\mu_{\rm max}$ diagram. 

\subsection{Including W1 and W2 from \textit{unWISE}}
\label{sec:unwise}

The \textit{unWISE} catalogue \citep{schlafly19} contains more than 2 billion sources all over the sky in the 3.6 and 4.5$\mu m$ (W1 e W2) bands from the analysis of the \textit{unWISE} coadds of the \textit{WISE} images\footnote{\url{http://catalog.unwise.me/}}. The \textit{unWISE} coaddition is described in \citet{lang14} and \citet{meisner17a, meisner17b, meisner18a}.

Concerning its predecessor \textit{ALLWISE}, this catalogue presents deeper images, since it involves the coaddition of all publicly available 3--5 microns \textit{WISE} imaging. This procedure is equivalent to increasing the total exposure time by a factor of 5 and allows the detection of magnitudes $\sim 0.7$ fainter (at a 5$\sigma$ level). This doubles the number of detections between redshifts 0 and 1, and triples between 1 and 2, totalizing more than half a billion galaxies. Another advantage of this new catalogue is the improvement in the modeling of the crowded regions with the \texttt{crowdsource}\footnote{\url{https://github.com/schlafly/crowdsource}} analysis pipeline \citep{schlafly18}, which optimizes the positions, fluxes, and the background sky to minimize the differences between observation and model. 

Our images and \textit{COSMOS2015} area are inside the \textit{unWISE} coadds, which makes it possible for us to do a cross-match between catalogues to include the near-IR information in our analysis.

The accuracy of photometric redshifts depends strongly on the number of available photometric bands. Our observations were done in only three bands per field. For this reason, we included in our sample (as well as in the \textit{COSMOS2015} catalogue) the \textit{W1} and \textit{W2} bands from the \textit{unWISE}.

The field of view of each of our fields correspond to two or four images (coadds) of the \textit{WISE} survey, and we have used \textit{topcat} \citep{taylor05} to perform the matching of the objects detected in \textit{unWISE} \textit{W1} and \textit{W2} bands with our catalogues. In this process, we removed the coadds overlapping edges to avoid object repetition, concatenated the data of each image, and, finally, made a cross-match between the resulting \textit{unWISE} catalogue and ours with a projected distance of $\leq 2.75$ arcsec. It is also important to point out that, in all fields, including \textit{COSMOS}, about 60\% of the cross-matched galaxies have measurement only in W1 or W2. These galaxies with missing photometry reduce the quality of the estimated photometric redshifts (Section \ref{sec:photo-zs}) by 1--2\% since less information is being used as input data in the machine learning process. However, given the expressive amount of galaxies with missing near-IR photometry, we decided to keep them for further analysis. We also checked the galaxies without photometry in the optical filters. They represent $\sim$1\% of the total sample of objects measured with SExtractor, and, after the cross-match with the \textit{unWISE} galaxies, only $\sim$0.02\% remains. Since this ratio is very low, and most of these galaxies already do not have measurements in one of the near-IR filters, we discarded them.

The flux measurements of \textit{unWISE} are in VEGA nanomaggies units \citep{finkbeiner04}. Therefore, we use the expressions provided by \citet{schlafly19} to convert them into AB magnitudes:\\

$m_{\rm {W1,AB}} = m_{\rm{W1,Vega}} + 2.699$

$m_{\rm{W2,AB}} = m_{\rm{W2,Vega}} + 3.339$ \\



\section{Analysis}
\label{sec:analysis}

At this stage, we have galaxy catalogues with photometric measurements in five bands (3 in the optical, from the CFHT images; and 2 in the near-IR, from unWISE). In this section, we discuss, initially, the estimation of photometric redshifts for each field,  the procedure adopted to evaluate the galaxy density field in a redshift interval containing each one of our triplets, and the comparison with the mock data. 

\subsection{Photometric redshifts} \label{sec:photo-zs}

There are basically two ways to calculate photo-zs: the \textit{template fitting} approach \citep{benitez00, arnouts02, ilbert06, tanaka15} and machine learning methods \citep[MLMs: e.g.,  ][]{collister04, almosallam16, sadeh16}. The first one relies on empirical \citep{coleman80} or synthetic spectra \citep{bruzual03, maraston05, charlot07} of different types of galaxies that are processed along with the photometric information of the observations, taking into account the telescope response and the filters characteristics. MLMs work with \textit{training data sets} comprised of objects with known redshifts to derive a relationship between the photometric measurements and the photo-zs. For this reason, they do not require physically motivated models, which helps to incorporate new observables into the inference and mitigates systematic errors \citep{sadeh16}.

We use \texttt{ANNz2}\footnote{\url{https://github.com/IftachSadeh/ANNZ}} \citep{sadeh16} to estimate photo-zs. This is a new implementation of the \texttt{ANNz1} package \citep{collister04}. \texttt{ANNz2} uses the \texttt{ROOT C++} software framework \citep{brun97}, that contains the \textit{Toolkit for Multivariate Data Analysis (TMVA)} package \citep{hoecker07}, allowing the choice of different algorithms to train MLMs. For this reason, unlike its predecessor, \texttt{ANNz2} incorporates different MLMs in its code (e.g., \textit{artificial neural network} (ANN), \textit{boosted decision trees} (BDT), among others), providing larger  versatility in the photo-zs inference, as well as probabilistic errors estimation. Like all MLMs, a sample with spectroscopic redshifts (spec-zs) is required and is split into \textit{training}, \textit{validation} and \textit{test sub-samples}. During training, the \textit{validation sub-sample} is used step-by-step to check the convergence of the solutions and evaluate the mapping between photometry and redshifts. The \textit{test sub-sample} -- which is not part of the training process -- allows an independent evaluation of the performance of the algorithm.

In addition, it is possible to operate \texttt{ANNz2} in \textit{single} or \textit{randomized regression}. The first one is the simplest configuration and generates a similar nominal product as \texttt{ANNz1}. However, \citet{sadeh16} shows that \texttt{ANNz2} has slightly superior performance. They also point out that the method used to calculate uncertainties has been significantly improved. The original version uses the propagation of input uncertainties to obtain those for the estimated redshifts, through the chain rule; \texttt{ANNz2} uses a data-driven method where it takes into account that objects with similar photometric properties should also have similar uncertainties in the photo-zs. For this, \texttt{ANNz2} uses the \textit{k-nearest neighbours} (KNN) method (see \citealt{oyaizu08}). On the other hand, \textit{randomized regression} uses several combinations of MLMs to compute photo-zs probabilistic distribution functions (PDF) for each galaxy. Then, these combinations are ranked following some metrics, like \textit{bias}, \textit{outlier fraction} and \textit{scatter}.

Another challenge for the current work is to find a homogeneous sample with a large enough number of galaxies up to spec-zs $\sim$ 1.5. High-z surveys with a large number of objects are in general biased in specific redshift slices, depending on the survey objectives, resulting in inhomogeneities that can bias their use as a training set in MLMs. In this work, we have used the \textit{COSMOS2015} catalogue, since it has photo-zs with high precision and homogeneous distribution up to $z\sim1.5$, as described in section \ref{sec:cosmos2015}. The training was done with \texttt{BDT} in the \textit{single regression} mode, since the randomized and \texttt{ANN} algorithms have a larger computational cost and the results are very similar when compared through the $Bias = <z_{\rm{in}} - z_{\rm out}>$, and the \textit{normalized median absolute deviation $\sigma_{\rm NMAD}$} metrics, proposed by \citet{molino17}, which evaluates the accuracy of the estimated redshift ($z_{\rm out}$) with respect to the \textit{input} ($z_{\rm in}$) value:

\begin{equation} 
     \sigma_{\rm NMAD} = 1.48 \times MEDIAN\left( \frac{(|\Delta z - MEDIAN(\Delta z)|)}{1 + z_{\rm in}}  \right),
     \label{eq:nmad}
\end{equation}
\noindent
where, $\Delta z = z_{\rm out} - z_{\rm in}$.

To obtain an unbiased sample when running \texttt{ANNz2} it is also important that the training, validation, and test samples (divided as 70\%, 15\%, and 15\%, respectively) be consistent between them and with our sample. Since the \textit{COSMOS2015} catalogue is deeper than our optical observations, we selected only galaxies inside the same interval of magnitudes applied at our images (see Table \ref{tab:phys_sample}). We compared the magnitude measurements of the two samples and found that, considering the whole magnitude interval of each field (see Table \ref{tab:phys_sample}), there is no statistical difference at 1$\sigma$ level between them and \textit{COSMOS2015}. Considering only the range of redshift of our final sample of galaxies at the triplet’s redshift slab (for slab criterion, see Section 3.2), there is no statistical difference at 1$\sigma$ level for all bands in the fields S2, S4, and S6. For the others, the difference is larger, and only at a 2$\sigma$ level, they can be considered statistically equal. We considered these results fair enough to proceed with the training process. 

\begin{figure*}
    \centering
    \includegraphics[width=\textwidth]{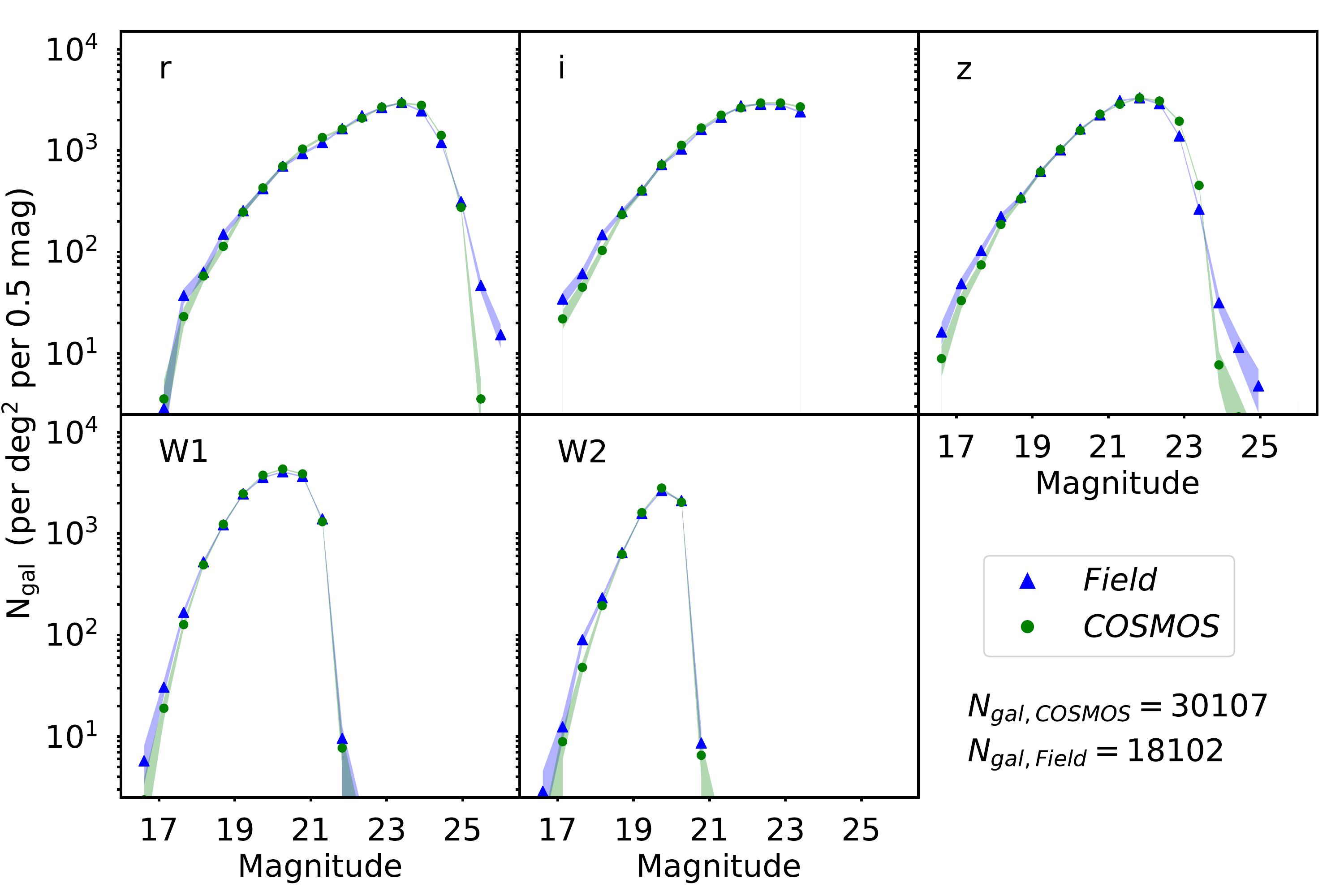}
    \caption{\small Galaxy counts for the S1 field (blue) and \textit{COSMOS} (green), after applying to both samples the magnitude limits quoted in Table \ref{tab:phys_sample}}.
    \label{fig:s1_counts}
\end{figure*}

We present the magnitude distributions in the 5 available photometric bands for the \textit{S1} field and the \textit{COSMOS2015} catalogue in Figure \ref{fig:s1_counts}. We also tested including cuts in other bands to better converge the faint-end between the two counts. However, we decided to keep cuts only in the reference band to avoid bias, and also because the difference in the results is not large and does not affect the final result qualitatively. For the other field distributions, see Appendix \ref{sec:apB}. We notice that there is an excess of bright galaxies in the S6 field ($i \lesssim 19$) compared to the COSMOS2015 sample. We ascribe this feature to a real excess of bright objects. In any case, galaxies that might be members of the triplet (see Figure \ref{fig:colour}) are expected to be significantly fainter ($i \gtrsim 21$), and are in a magnitude interval where our galaxy counts are consistent with those of COSMOS2015.

\begin{figure}
    \centering
    \includegraphics[width=\columnwidth]{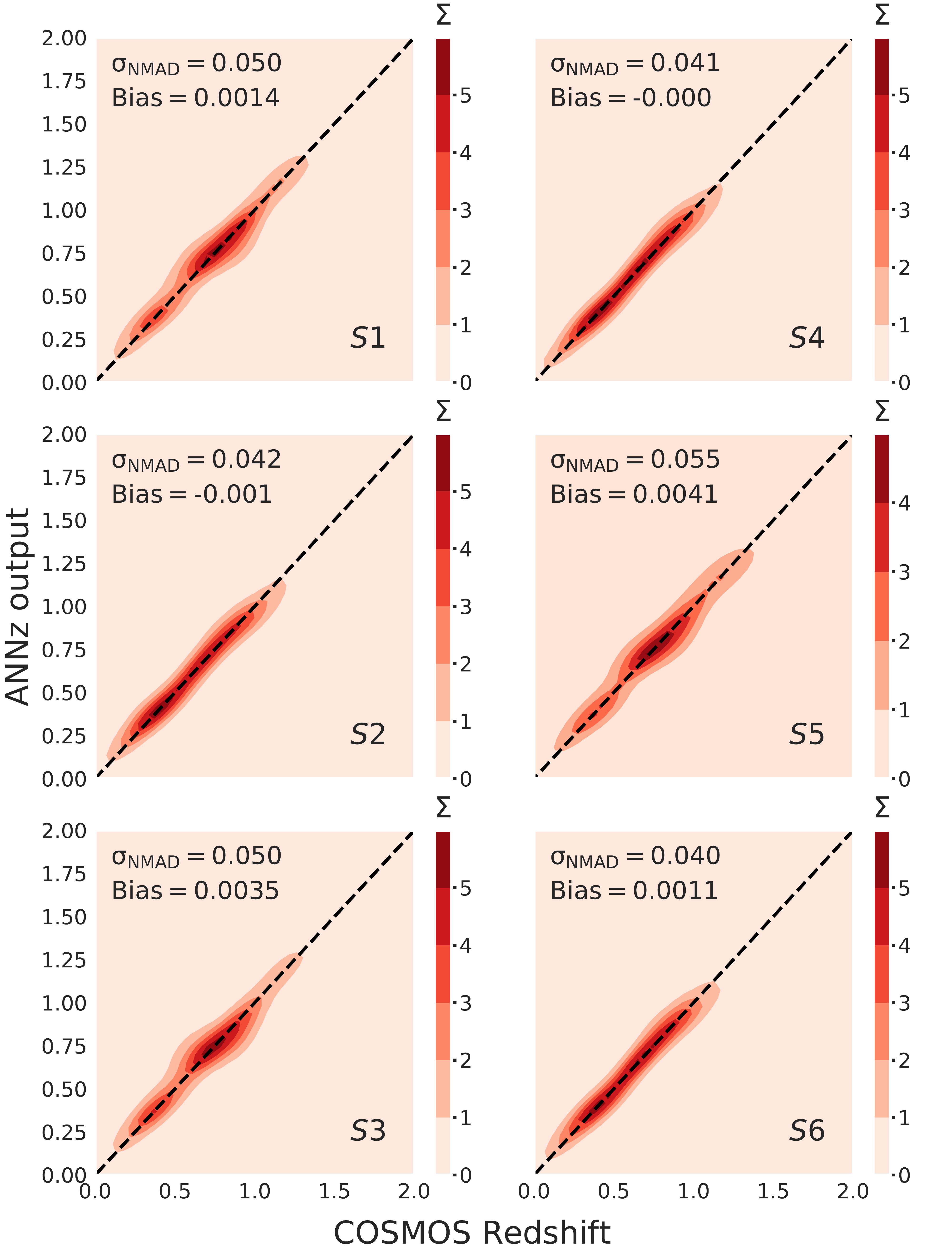}
    \caption{\small Photometric redshift estimation using \texttt{ANNz2} for the validation and test data sets. The contours represent the density distribution calculated with a gaussian KDE function (see Equation \ref{eq:kde}). The $x = y$ relation is denoted by the dashed line. The higher the density around the $x = y$ line, the lower will be the Bias and the $\sigma_{NMAD}$, indicating higher precision and accuracy, respectively.}
    \label{fig:annz_perf}
\end{figure}

\begin{figure}
    \centering
    \includegraphics[width=\columnwidth, trim= 0 3.5cm 0 1.8cm]{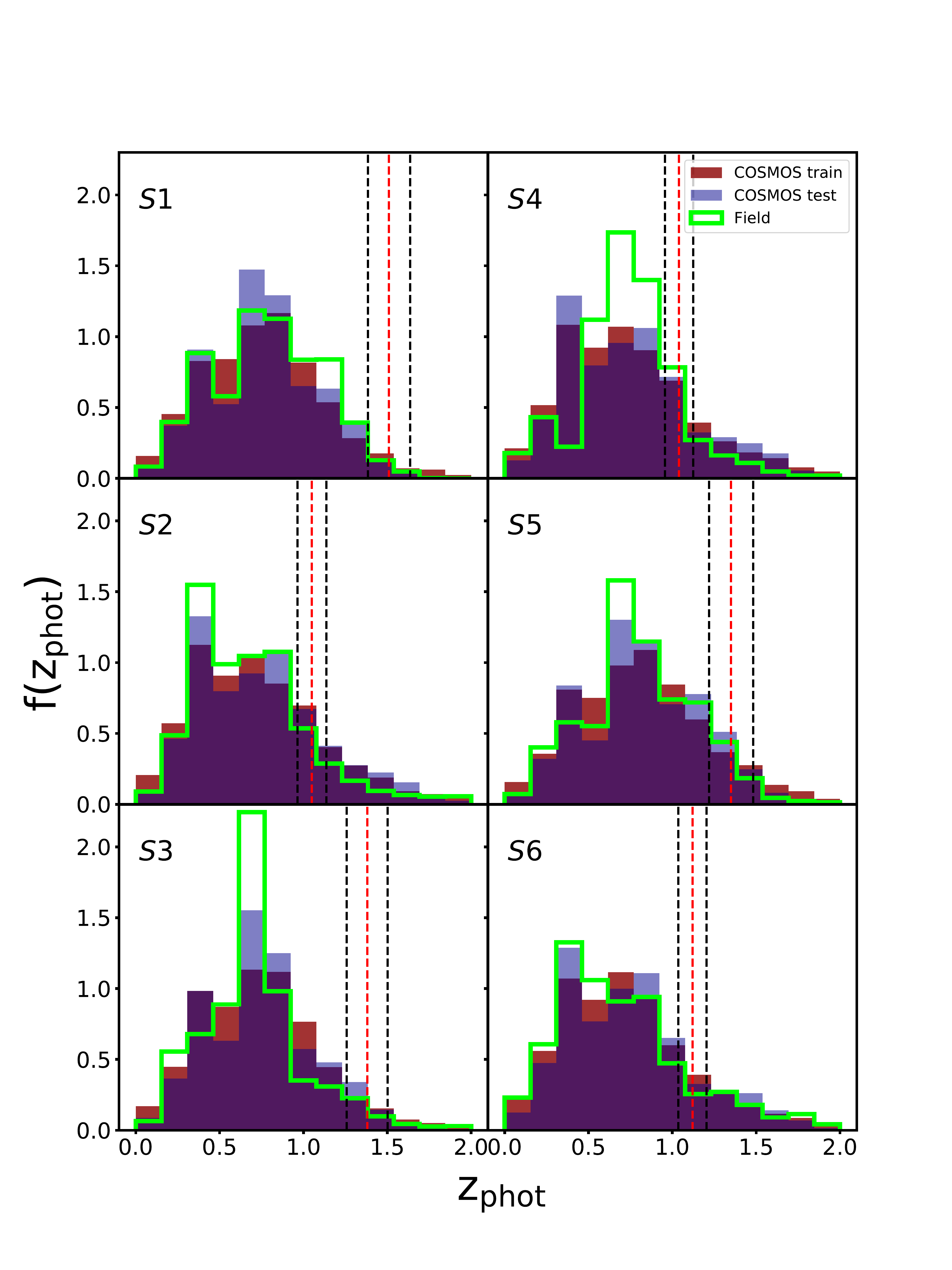}
    \caption{\small Photometric redshift distribution for COSMOS test-set galaxies (blue), COSMOS training-set galaxies (dark red), and for the CFHT fields (lime green), adopting the same magnitude limits. The red and black vertical dashed lines stand for the $\bar z$ and $\bar z_{\rm slab}$, respectively.}
    \label{fig:z_dist}
\end{figure}

We tested different input configurations (only colours between adjacent bands; adding the reference band; and adding more bands with different combinations) comparing which one gives the best results based on the metrics described above. The best results were obtained with all magnitude measurements and colours: \textit{r}, \textit{i}, \textit{z}, \textit{W1}, \textit{W2}, \textit{r-i}, \textit{i-z}, \textit{z-W1}, \textit{W1-W2}, for the S1, S4, and S6 fields; and \textit{g}, \textit{r}, \textit{i}, \textit{W1}, \textit{W2}, \textit{g-r}, \textit{r-i}, \textit{i-W1}, \textit{W1-W2}, for S2, S3, and S5. To check the performance of the algorithm, we plot $z_{\rm in}$  versus $z_{\rm out}$ for the test sub-samples, as well as the $Bias$ and the $\sigma_{\rm NMAD}$ values in Figure \ref{fig:annz_perf}. Figure \ref{fig:z_dist} presents the redshift distributions of each field as well as for the  COSMOS training and test samples.

For all cases in Figure \ref{fig:z_dist}, the distributions between the COSMOS training and test samples follow similar shapes, which is another indication (in addition to the photo-z metrics) that the network training process was successful. The photometric redshift distributions of our fields present some minor discrepancies compared to the COSMOS samples, probably due to cosmic variance, but these differences are negligible in the bins corresponding to the triplets' redshifts.

It is also important to say that we also tested the estimates without and with the galaxies that have no measurements in W1 or W2 (as described in \ref{sec:unwise}), obtaining $\sigma_{\rm NMAD}$ ranges of 2.4--3.8\% and 4.0--5.5\%, respectively. Galaxies that have no measurements in one of these two bands represent 60\% of the sample, so we decided to keep them even if this represents a cost in the performance of \texttt{ANNz2}. These values are obviously much higher than the accuracy achieved by the \textit{COSMOS2015} photo-z estimation, since the amount of information (inputs) available is much less. Still, our $\sigma_{\rm NMAD}$ values are comparable with those obtained in other works \citep[e.g.,][]{vanderburg20, jian2020}.

\subsection{The density field}
\label{sec:density_field}

To verify how galaxies are spatially distributed in a certain field at the triplet redshifts, we computed overdensity significance maps using the \textit{kernel density estimator} \cite[KDE; e.g.,][]{ivezic14}. The galaxy surface density at a point $x$ is given by
\begin{equation} 
    \Sigma(x) = \frac{1}{Nh^{2}}\sum_{\rm i=1}^{N} K\left(\frac{d(x,x_{\rm i})}{h}\right),
     \label{eq:kde}
\end{equation}
where $x_i$ represents the coordinates of one of the $N$ galaxies with photometric redshifts, $K(u)$ is the kernel function (assumed Gaussian), $d(x,x_{\rm i})$ is the projected (Euclidian) distance between $x$ and $x_{\rm i}$, and $h$ is the kernel bandwidth. 

The bandwidth can be chosen with statistical or physical criteria. In the former case it is common to adopt thumb rules (e.g. \textit{Scott's rule}) or \textit{Cross-validation}. We have noticed, however, that both cases lead to large values of $h$ and, consequently, to too smooth maps of the galaxy distribution. A more physical approach is to adopt a sensible scale for the kernel bandwidth. \cite{chiang13} have used the Millennium simulation to estimate the effective radius ($R_{\rm e}$) of $z=0$ galaxy clusters progenitors as a function of redshift, finding that, at z $\sim$ 1.3, the typical diameter ranges from $4$ to $15$ cMpc (lower and upper limit of Fornax-type and Coma-type progenitors, respectively). Motivated by this result, we decided to adopt $h=5$ and $10$ cMpc, to perform our analysis. The smaller bandwidth is better to probe $\sim$group scale structures, while the larger one is more appropriate for the more massive objects. For our data, values lower than $5$ cMpc are not convenient, since the average distance to the fifth nearest neighbor is $\sim 4.5$ cMpc. In other words, below this value, the overdensity significance maps are very fragmented into small sets of galaxies, deteriorating the map's information quality due to shot noise.

Another interesting parameter that can be used with the kernel function is a weight ascribed to each galaxy. We use the redshift estimates and their respective errors to calculate the probability that each galaxy is in the QSOs redshift slab. For this, we assume that a photo-z and its correspondent error are the mean and standard deviation of a Gaussian probability density function. Then, we calculate the probability $p$ that a galaxy is in a redshift slab width $\bar z_{\rm slab} = [\bar z - \sigma_{\rm NMAD}(1 + \bar z)$, $\bar z + \sigma_{\rm NMAD}(1 + \bar z)]$ (see Table \ref{tab:phys_sample}). We use $p$ as the weight of the galaxy in the map. That is, the closer, in redshift, the galaxy is to the QSOs system, the greater is its contribution to the density map. The definition of the weight for a given galaxy in our maps is
\begin{equation}
    w_{\rm i}(x_{\rm i},y_{\rm i}) = \frac{p_{\rm i}} {\sum_{\rm {j=1}}^{\rm N} p_{\rm j}}
\end{equation}
where $x_{\rm i}$ and $y_{\rm i}$ are the coordinates of a given galaxy, and $p_{\rm j}$ is the probability that the $j$-th galaxy is in the redshift slab of the triplet. 

Each field was then covered by a $100\times100$ grid (where each pixel has a width of $\sim 0.6$ arcmin, the kernel surface density  was computed at each node of the grid, and, finally, we determined the overdensity significance as

\begin{equation} 
    \sigma_{\rm gal} = \frac{\Sigma - \bar \Sigma} {\sigma},
     \label{eq:contrast}
\end{equation}
where $\bar \Sigma$ and $\sigma$ are the mean and standard deviation of the KDE density for the values in the grid. 

\begin{figure*}
\centering
\includegraphics[width=\textwidth]{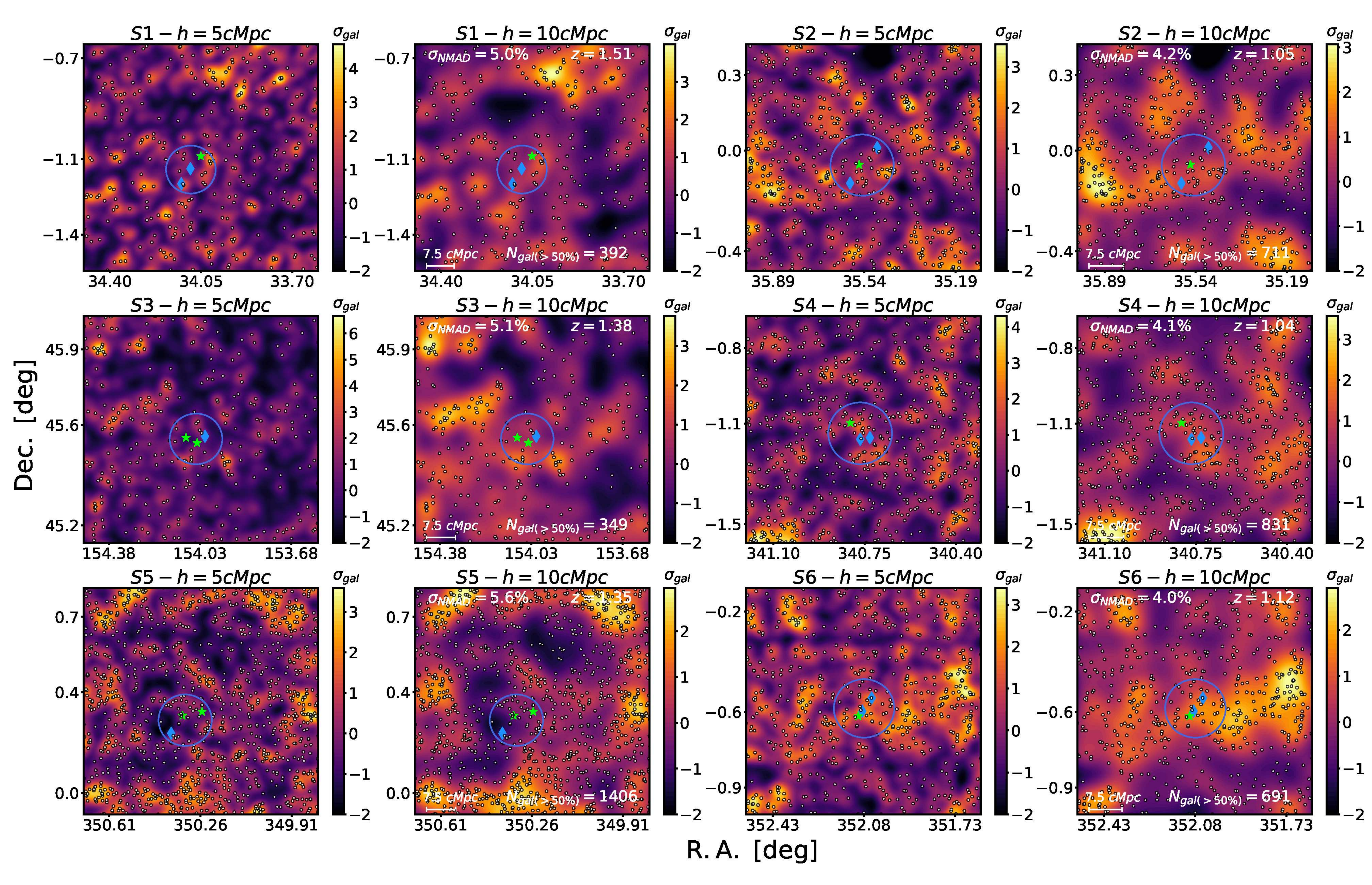}
\caption{\small significance overdensity maps for CFHT/Megacam observations. Green stars mark the position of RL QSOs, while blue diamonds the RQs; white dots are the galaxies with $p > 50\%$; blue circles have 7.5 cMpc projected radius and is centered in the QSOs system's centroid; The title of each map shows the field and the correspondent bandwidth; In each field for $h = 10$ cMpc map, we also present general information as the $\sigma_{\rm NMAD}$ and the number of galaxies with p $\geq$ 50\%.}
\label{fig:o_maps}
\end{figure*}

\begin{figure*}
\centering
\includegraphics[width=\textwidth]{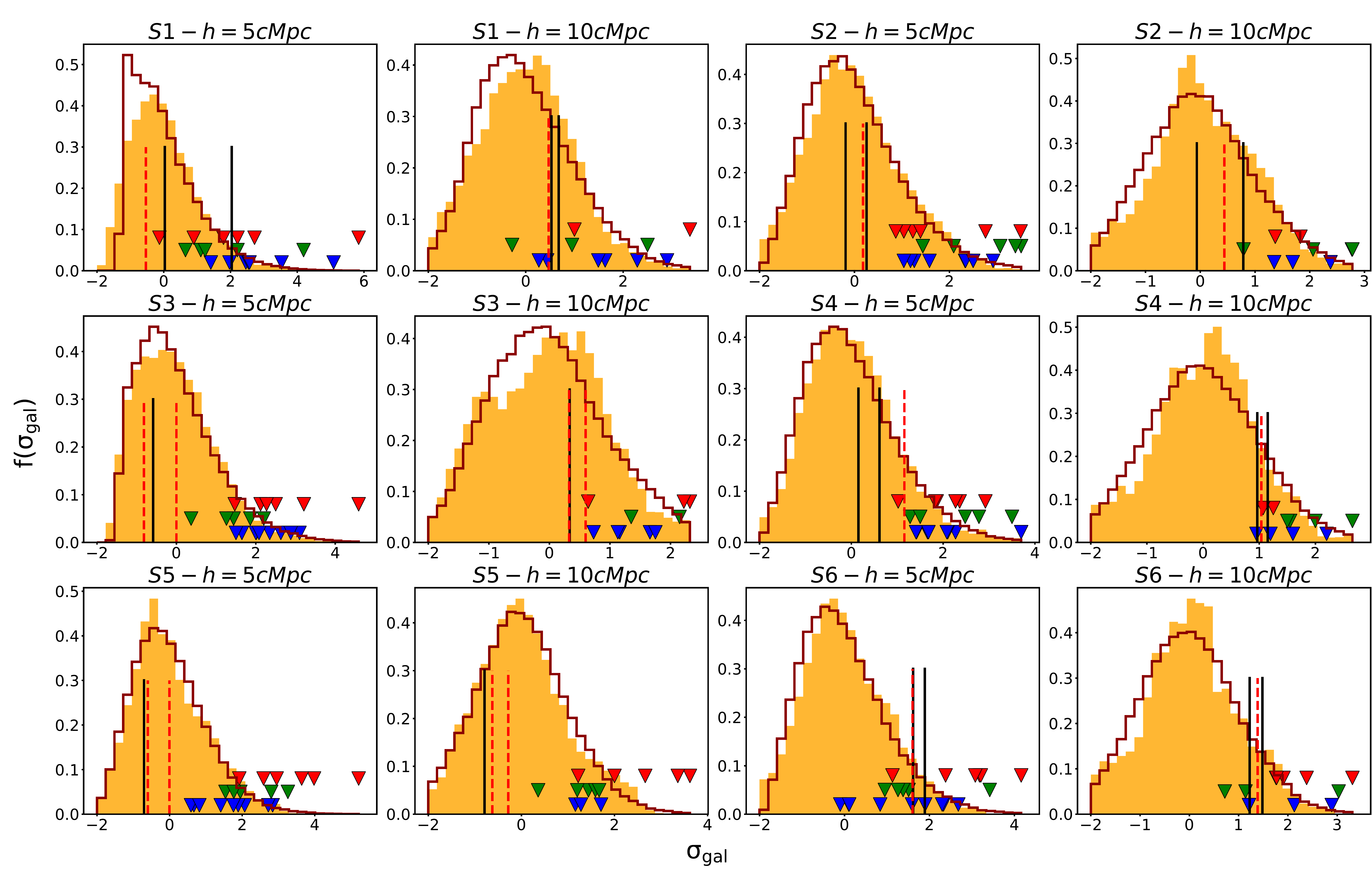}
\caption{\small The orange histograms are the overdensity significance distributions. The red dashed and the solid black vertical lines stand for the measurement of the overdensity significance where RL QSO or RQ QSO members of the triplet are (see Table \ref{tab:measurements}), respectively. The red line distributions are from the simulated sample. Red, green, and blue triangles mark the significance overdensity, $\sigma_{\rm gal}$, of Coma, Virgo, and Fornax type protoclusters, respectively. }
\label{fig:contrast_hist}
\end{figure*}

We present in Table \ref{tab:measurements} the values of $\sigma_{\rm gal}$ associated to each of the QSOs of our sample for the two different bandwidths chosen for our analysis. Figure \ref{fig:o_maps} shows the resulting overdensity significance maps. In Figure \ref{fig:contrast_hist}, we present the $\sigma_{\rm gal}$ distributions for galaxies with $p$ $\geq$ 50\%, indicating the densities associated to the triplet's members. This figure shows that the triplets avoid the densest regions of each field. However, the overdensities associated with the quasars in the S4 and S6 fields are larger than one and consistent with protoclusters detected in our mock analysis (see Section \ref{sec:comp_mocks}). Yet, this result alone is not enough to consider these two triplets as protoclusters candidates. Indeed, as shown in the next section, they probably are not. 

Also, since the fields have a large area, to avoid the case where there is a huge sheet-like structure that is not being detected because the entire redshift slab of the triplet is overdense, we compared them to the neighbouring slabs, and noticed that the former are not overdense compared to the latter in all fields.

\begin{table*}
\caption{\small Measurements of the triplet members overdensity significance in the density field. The first two columns are the field ID and the QSO members of each triplet. The next three columns are related with the h = 5 cMpc maps: the overdensity significance associated with each QSO and the mean surface density and its standard deviation for all pixels in the map (in cMpc$^{-2}$). The next three columns contain the same information for the h = 10 cMpc maps. The  last two columns are the medians of the r - i colour for the field galaxies and for the galaxies closer to the QSO. For more details,  see Sections \ref{sec:density_field} and \ref{sec:discussion}.}
\begin{tabular}{cccccccccc}
\hline
Field  & QSOs  & $\sigma_{\rm{gal, 5}}$ & $\bar \Sigma_{\rm 5}$ (cMpc$^{-2}$) & $\sigma_{\rm 5}$  (cMpc$^{-2}$) & $\sigma_{\rm{gal, 10}}$ & $\bar \Sigma_{\rm 10}$ (cMpc$^{-2}$) & $\sigma_{\rm 10}$ (cMpc$^{-2}$) &  
\multicolumn{1}{l}{Med$(r-i)_{\rm field}$} & \multicolumn{1}{l}{Med$(r-i)_{\rm trip}$} \\
\hline
\multirow{3}{*}{S1} & SDSSJ021612-010519   & -0.536 & \multirow{3}{*}{0.966} & \multirow{3}{*}{0.520} & 0.469 & \multirow{3}{*}{0.937}&\multirow{3}{*}{0.294}&\multirow{3}{*}{0.479}& \multirow{3}{*}{0.462}                        \\
                    & SDSSJ021622-010818* & 0.033 & & & 0.529          &                        &                        &                                          &
                    \\
                    & SDSSJ021630-011154* & 2.034 & & & 0.676          &                        &                        &                                           &                                                \\
\hline
\multirow{3}{*}{S2} & SDSSJ022158+000043* & 0.179 & \multirow{3}{*}{0.963} & \multirow{3}{*}{0.436} & 0.440& \multirow{3}{*}{0.930} & \multirow{3}{*}{0.285} & \multirow{3}{*}{0.690}& \multirow{3}{*}{0.692}                        \\
                    & SDSSJ022214-000322  & 0.254 & & & -0.063         &                        &                        &                                             &                                               \\
                    & SDSSJ022223-000745* & -0.187 & & & 0.785          &                        &                        &                                             &                                               \\
\hline
\multirow{3}{*}{S3} & SDSSJ101603+453316* & 0 & \multirow{3}{*}{0.963} & \multirow{3}{*}{0.567} & 0.599          & \multirow{3}{*}{0.934} & \multirow{3}{*}{0.349} & \multirow{3}{*}{0.447}                      & \multirow{3}{*}{0.454}                        \\
                    & SDSSJ101610+453142   & -0.819 & & &0.326          &                        &                        &                                             &                                               \\
                    & SDSSJ101620+453257   & -0.585 & & &0.335          &                        &                        &                                             &                                               \\
\hline
\multirow{3}{*}{S4} & SDSSJ224255-010924* & 0.153 & \multirow{3}{*}{0.962} & \multirow{3}{*}{0.396} & 1.151          & \multirow{3}{*}{0.928} & \multirow{3}{*}{0.264} & \multirow{3}{*}{0.784}                      & \multirow{3}{*}{0.888}                        \\
                    & SDSSJ224304-010946* & 0.612 & & & 0.964        &                        &                        &                                             &                                               \\
                    & SDSSJ224313-010552   & 1.155 & & & 1.037        &                        &                        &                                             &                                               \\
\hline
\multirow{3}{*}{S5} & SDSSJ232101+001923   & -0.600 & \multirow{3}{*}{0.961} & \multirow{3}{*}{0.390} &-0.282         & \multirow{3}{*}{0.934} & \multirow{3}{*}{0.267} & \multirow{3}{*}{0.545}                      & \multirow{3}{*}{0.565}                        \\
                    & SDSSJ232119+001826   & -0.003 & & &-0.624         &                        &                        &                                             &                                               \\
                    & SDSSJ232129+001413* & -0.705 & & &-0.791         &                        &                        &                                             &                                               \\
\hline
\multirow{3}{*}{S6} & SDSSJ232812-003238* & 1.617 & \multirow{3}{*}{0.964} & \multirow{3}{*}{0.366} & 1.225          & \multirow{3}{*}{0.938} & \multirow{3}{*}{0.269} & \multirow{3}{*}{0.643}                      & \multirow{3}{*}{0.691}                        \\
                    & SDSSJ232821-003547* & 1.891 & & & 1.484          &                        &                        &                                             &                                               \\
                    & SDSSJ232824.5-003658 & 1.603 & & & 1.385          &                        &                        &                                             &    
                    \\
\hline
\end{tabular}%
\label{tab:measurements}
\end{table*}

\subsection{Color-magnitude diagrams}
\label{sec:colour}

Now we discuss the properties of the galaxy population around the triplets with colour-magnitude diagrams.

\begin{figure*}
\centering
\includegraphics[width=\textwidth]{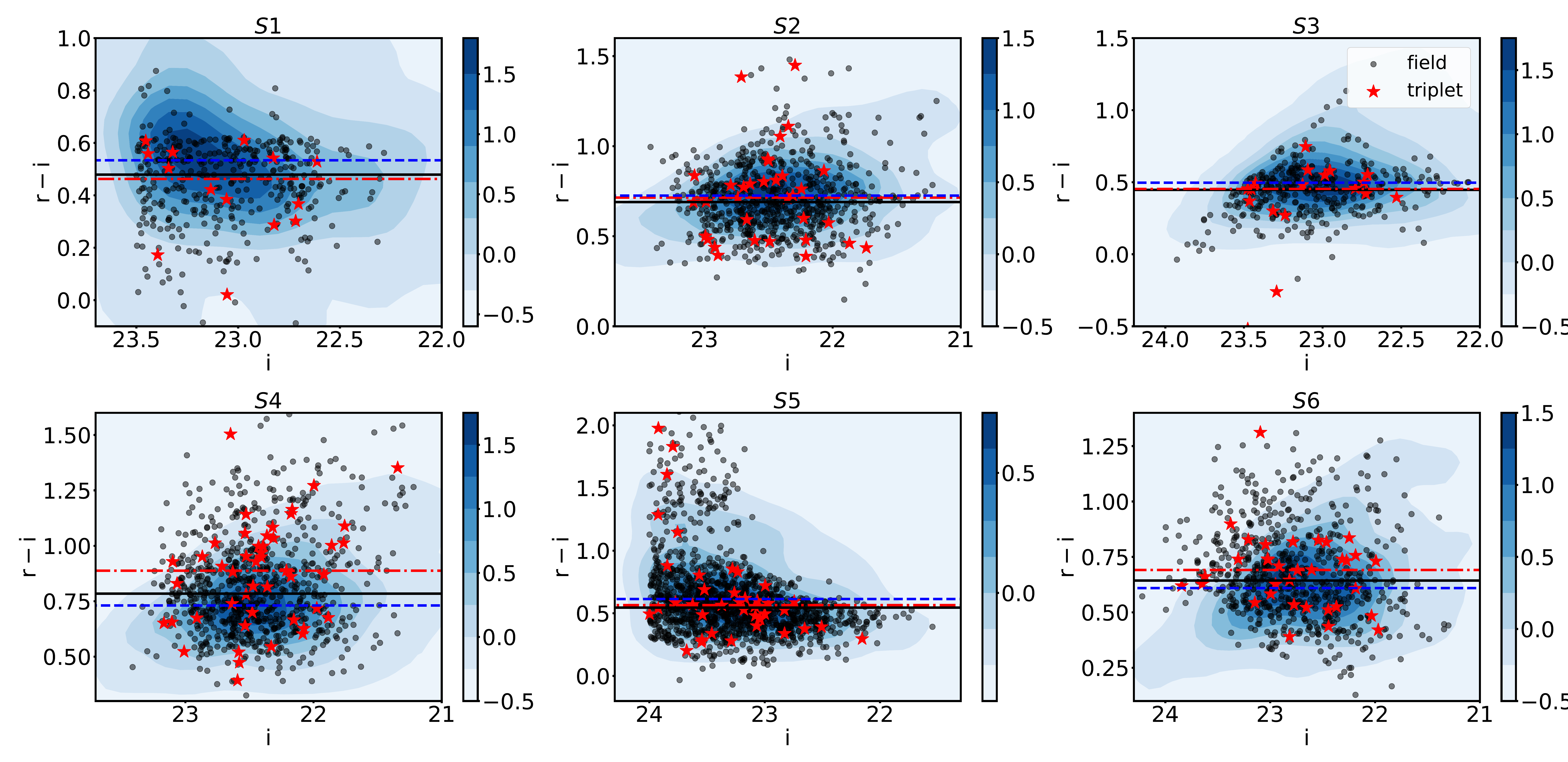}
\caption{\small Colour-magnitude diagrams for our six fields and for the COSMOS2015 training set galaxies. Black points are galaxies within the slab, whereas red points are the galaxies in the encircled regions; red dashed and black solid lines are the $r-i$ median for the galaxies inside and outside the circle, respectively. The blue contours are the overdensity significance calculated from a Gaussian KDE $\Sigma(i,r-i)$, of the COSMOS2015 training set galaxies, and the blue horizontal line is the median of its $r-i$ colour.}
\label{fig:colour}
\end{figure*}

Figure \ref{fig:colour} shows $(r-i)$ vs $i$ diagrams with galaxies ``inside'' and ``outside'' the encircled regions in the density field\footnote{The circle has a 7.5 cMpc projected radius, while the distance between the centroid of the quasars triplet, and the coordinates of each galaxy, are calculated with the angular distance equation.} shown in Figure \ref{fig:o_maps}, as well as for the COSMOS2015 galaxies in the same redshift interval, for a qualitative comparison of the distributions.  Figure \ref{fig:colour} indicates that galaxies closer to the triplet in the S4 field tend to be redder in the $r-i$ colour than the others. This trend is less strong for the other triplets. The median colours involved in this analysis are also presented in Table \ref{tab:measurements}. 

To be more quantitative, we have compared through a two-dimensional Kolmogorov-Smirnov test the colour distribution of galaxies within and outside the blue circles of the six fields in Figure \ref{fig:o_maps}. The resulting p-values that the inside and outside samples were drawn from the same distribution are 0.91, 0.27, 0.72, 0.04, 0.85, and 0.40, from S1 to S6. The null hypothesis can be rejected for S4 at a confidence level of 96\%, contrary to the other fields.

\begin{figure}
\includegraphics[width=\columnwidth]{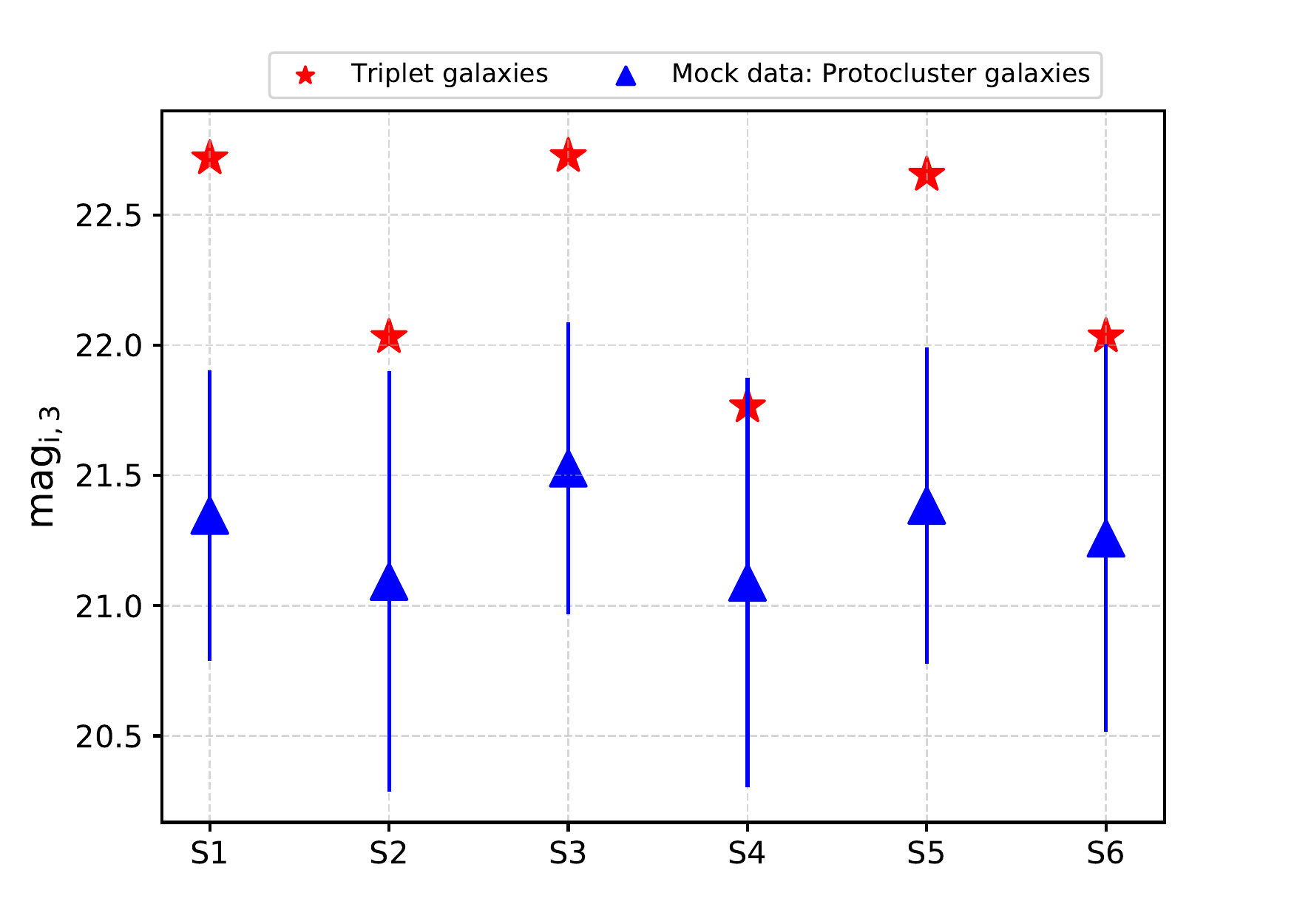}
\caption{\small The red stars denote the $i$ magnitude of the third brightest galaxy of the triplet galaxies, while blue triangles stand for the mean $i$ magnitude of the third brightest galaxies of the protoclusters in our mocks.}
\label{fig:m3}
\end{figure}

We can obtain additional information on the nature of the triplets environments from the i-band magnitude distribution of their galaxies (represented by the red stars in Figure \ref{fig:colour}) in these colour-magnitude diagrams. They indicate a paucity of bright galaxies in the triplets, compared to the overall magnitude distribution in the slab. We show in Figure \ref{fig:m3} the mean value of the third brightest-galaxy magnitude in our mock protoclusters and in the triplets. It is possible to observe that the former are consistently brighter than the latter in most cases. The i-band magnitude differences are 1.37, 0.94, 1.20, 0.68, 1.27, and 0.77, from S1 to S6. the smallest difference is found in the S4 field, which is the only case in which there is no statistical difference within one $\sigma$.

\subsection{Comparison with simulations}
\label{sec:comp_mocks}

We have reproduced our method to estimate overdensity significance maps with our mock samples. Both observed and simulated datasets present some differences in the number of galaxies due to the cross-match of our imaging data with the unWISE catalog, otherwise, they would be similar. Additionally, the magnitude estimation for the PCcones did not consider dust emission, which becomes non-negligible in the MIR, in particular in the \textit{W2} band. Hence, we cannot estimate photometric redshifts directly from the mock magnitudes. However, we statistically address these differences, as explained below.

\subsubsection{Photometric redshifts for mock galaxies}

We generated mock photo-zs according to the redshift of the galaxies in the PCcones, following the procedure used by \citet{krefting20}. The mock photo-z for a galaxy at $z_i$ and reference band magnitude $m_{\rm ref,i}$ is drawn from a Gaussian with mean $z_i$ and standard deviation equals to the typical error in the photo-z measurements\footnote{\citet{krefting20} uses as standard deviation the typical width of the photometric redshift probability density distribution, $p(z)$, which are related with the photometric redshift errors.} for galaxies with magnitude $m_{\rm ref,i}$. We have also added catastrophic redshifts as described by \citet{krefting20}. Assuming a representative outlier fraction (9.5, 7.5, 10.7, 6.5, 11.4, and 6.7\% for each field, respectively) these mock catastrophic photo-zs were drawn uniformly between $0 \leq z \leq 3.0$. With this technique, we achieved a $\sigma_{\rm NMAD}$ for the simulated photo-zs within $\sim 0.5$\% from those obtained for each field.

\subsubsection{Overdensity significance maps}

Our simulated sample presents an excess of galaxies compared to the observational sample, and we need to remove this excess to perform a reliable comparison. We made it by selecting all galaxies in the mock with photo-zs within the redshift slab. Then, we compute the overdensity significance maps from a subsample of galaxies according to the effective number of objects within the analyzed redshift interval. For a given slab, we define this quantity as the sum of all probabilities of the galaxies to be within it. These sums (extracted from the observational photo-zs estimates) are 595, 910, 711, 899, 1594, and 837 for each field, respectively. Notice that at this step all weights are the same. Additionally, we used the same 100$\times$100 pixels grid and the two different kernel bandwidth $h$, as we described above.

We present in Figure \ref{fig:contrast_hist} the overdensity significance distribution of our mock fields (dark red step histogram). Despite the applied method for the mock sample be not the same as that performed to observational data, in general, both observed and simulated distributions are similar, with exception of the S6, which shows a peculiar distribution. As we mention in Section \ref{sec:photo-zs}, the S6 field presents a large number of galaxies at higher redshifts.

\subsubsection{Cluster detectability}
\label{sec:detectability}

Following the same procedure described in \citet{araya20}, we first identify the overdensity peaks in each PCcone field, by comparing each pixel of these maps with the adjacent ones (in a $3 \times 3$ pixels matrix). If the central pixel has the highest value, we select its position as an overdensity peak. 
After, we link the single protocluster inserted into each PCcone with the overdensity peaks within a radius of 5 comoving-Mpc. If there is more than one peak in the linking area, then we attributed the denser one to the protocluster. 
We present the overdensity measurements of these structures for each field and the adopted bandwidth in Figure \ref{fig:contrast_hist}. The protoclusters that do not appear in the figure are all those without an overdensity peak within a projected distance of 5 cMpc of their centers. In general, using a kernel bandwidth of 5 cMpc, we have detected in our mocks 19, 19, 20, 19, 20, and 20 protoclusters out of the 20 simulated structures, respectively. For a bandwidth of 10 cMpc, the corresponding numbers are 11, 9, 10, 10, 13, and 11, respectively. Therefore, at least 97.5\% of the protoclusters are associated to an overdensity peak within a radius of 5 cMpc, if we use $h = 5$ cMpc, whereas this fraction is of only 53.3\% when we use $h=10$ cMpc. This occurs due to projection effects: with the high bandwidth value, other structures displaces the overdensity peak away from the position of the simulated protoclusters. Projection effects can also amplify the measured overdensities of structures and random regions, explaining why we detected some Fornax and Virgo-like progenitors in denser regions instead of Coma-like.


\section{Discussion}
\label{sec:discussion}

Some observations show that AGNs can reside in regions with a high density of galaxies in the universe \citep{seymour07}, and that is why they are often discussed as potential probes of high redshift structures. There are several of works along this line, and many structures have already been identified in this way. An example is the CARLA survey, which identified 200 cluster candidates at 1.3 < z < 3.2 from a survey of 420 radio-loud QSOs or radio-galaxies \citep{wylezalek13}. A spectroscopic follow-up with HST for the 20 densest candidates up to z = 2.8 confirmed 16 as high redshift clusters \citep{carla_survey}. However, the follow-up of more systems is required to establish the frequency of real clusters in this sample.

Also, several studies have focused on quasar/AGN systems, such as quasar pairs, with results somewhat contradictory. For instance, \cite{boris}, through the analysis of galaxy overdensities, their richness, and the identification of the red sequence in colour-magnitude diagrams, reported that 3 out of 4 pairs within the redshift interval $0.9 < z < 1$, presented strong evidence of being part of galaxy clusters. \cite{green11} studied the colour properties in the visible and possible X-ray emission of a hot IGM from 7 pairs extracted from SDSS-DR6 at $0.4 < z < 1$,  finding no evidence of association of these systems with galaxy clusters in their sample. Another interesting example is the work of \cite{onoue18}, who studied quasar pair environments at $z \sim 1$ and $z >$ 3, using optical catalogues from the HSC-SSP survey \citep[][DRS16A]{aihara17}. They found two pairs at $z =$ 3.3 and 3.6 associated with regions of high overdensity (5 $\sigma$ above the mean of the distribution), suggesting that pairs of luminous QSOs seem to be good tracers of protoclusters at these redshifts. At $z \sim$ 1, they worked with a total sample of 38 pairs and found that 19\% of them are in massive environments (> 4$\sigma$), concluding that pairs of QSOs at this redshift are also good tracers of matter-rich environments. 

Here we have extended this type of approach by analyzing QSO triplets at $1 < z < 1.5$, with the requirement that at least one member of the system is a radio-loud quasar. Our main results are depicted in Figure \ref{fig:o_maps} and \ref{fig:contrast_hist}, and Table \ref{tab:measurements}. The blue circles in the figure have a radius of 7.5 cMpc in projected distance. Our motivation for the adopted kernel bandwidth is related to the typical size of massive protoclusters, as well as to the number of galaxies in each field. From our simulations, we can recover a significant fraction (97.5\%) of protoclusters with a bandwidth of 5 cMpc.  For $h = 10$ cMpc, many low mass protoclusters do not reside in the densest regions in the field because the higher value of $h$ dilutes the overdensities of compact regions, such as groups and low-mass protoclusters. Additionally, irrespective of the bandwidth, the overdensities are ``washed-out'' due to the redshift uncertainties. The redshift slabs, which are defined according to the photometric redshift accuracy, are about five times wider than the redshift separation between the QSOs, as well about seven times wider than the typical $3\sigma$ velocity dispersion of Coma-type protoclusters at $1.0 \leq z \leq 1.5$ (1567 km/s). Nevertheless, we notice from Figure \ref{fig:contrast_hist} that overdensities associated with protoclusters are at the high-density end of all distributions for all the field emulations, independently of the kernel bandwidth. It implies that given our observational constraints and photometric redshift uncertainties, we still can separate real structures from random regions. 

Regarding the  overdensity significance, we find that, in general, the triplets tend to avoid overdense regions. However, for the S4 and the S6 fields this significance is consistent with that of protoclusters detected in our mock sample (see Table \ref{tab:measurements} and Figure \ref{fig:contrast_hist} for the overdensity significance associated with each QSO). At this point, it is reasonable to question whether the S4 and S6 overdensities are real or artifacts produced by projection effects.

In the case of S4, we notice that, for $h = 5$ cMpc, only the radio-loud quasar has an overdensity significance consistent with that of the mock protoclusters (see Figure \ref{fig:contrast_hist}). With $h = 10$ cMpc, this overdensity decreases a bit for the radio-loud quasar and grows for the other two, and the whole triplet acquires a significance consistent with the mock protoclusters. S4 has, also, a redder galaxy population close to the quasars (Section \ref{sec:colour} and Figure \ref{fig:colour}). On the other hand, the magnitude of its third brightest galaxies is smaller than expected in the mocks, suggesting  that the galaxy population close to the triplet is too faint. This indeed is observed for all triplets (Section \ref{sec:colour} and Figure \ref{fig:m3}), but S4 shows the lowest difference in the sample ($\Delta i = $0.68). A possible explanation is that this triplet inhabits an evolved group or a poor protocluster, with the radio-loud quasar at the cluster center as its Brightest Cluster Galaxy, since it is the brightest object in this redshift slab. 

The S6 triplet has an overdensity consistent with that expected from mock protoclusters. It is interesting to notice that its radio-loud quasar has the smallest black hole mass of our sample. We propose that, like S4, S6 is at most a poor protocluster because, as the other triplets, it lacks a population of bright galaxies. 

Summarizing, we conclude that none of the 6 fields shows  strong evidence for our triplets being members of massive objects, such as a rich high-z cluster or protocluster.The triplets in our sample, actually, tend to avoid the densest structures.


\section{Summary}
\label{sec:sum}

In this work, we have investigated the environment of six quasar triplets at 1 $\lesssim z \lesssim$ 1.5 making use of multiband CFHT/Megacam images, complemented with near-infrared photometry (3.6 and 4.5$\mu$m) from the \textit{unWISE} survey.

We have used photometric redshifts to identify galaxies at the redshifts of the triplets. These photo-zs were obtained with the \texttt{ANNz2} software trained with the accurate photometric redshifts of an enlarged version of the  \textit{COSMOS2015} catalogue, where we included the W1 and W2 \textit{unWISE} bands. This allowed us to obtain typical accuracies (measured with the $\sigma_{\rm NMAD}$ metrics) of 0.04 when compared with the spectroscopic training set.

The density field in a redshift slab of width $\Delta z = \sigma_{\rm NMAD} \times (1 + z)$ was obtained with a Gaussian Kernel Estimation with a bandwidth of 5 and 10 cMpc. The contribution of the galaxies for the map was weighted by the probability of the galaxy be in the redshift slab based on the photometric redshift estimate and its error.

We reproduce our method to compute overdensity significance maps for a set of lightcones, dubbed PCcones. These mocks were constructed by placing a desired structure at the redshift of the QSO triplets. In general, our results indicate that despite our observational constraints and photometric redshift uncertainties, we can separate satisfactorily real structures from random regions.

Our analysis shows that none of our six triplets present significant evidence of being part of a massive structure. In one case (S4), the overdensity significance and the colour-magnitude diagram suggest that the triplet might inhabit a group or a poor protocluster at $z = 1.04$.

\section*{Acknowledgements}
MCV acknowledges the support from São Paulo Research Foundation (CAPES/FAPESP agreement) through an MSc fellowship (grant \#2018/01469-6). PAA thanks the Conselho Nacional de Desenvolvimento Cient\'ifico  e Tecnol\'ogico (CNPq-Brasil) for supporting his MSc scholarship (133350/2018-5). LSJ acknowledges support from FAPESP (2017/23766-0) and CNPq (304819/2017-4). We thank Eduardo S. Cypriano for enlightening discussions and very useful suggestions in the course of this work. We also thank Erik V. R. de Lima for his assistance in the photometric redshifts section, and Dr. Scott Croom, who kindly make available the repository of the QSOs spectra images of the 2SLAQ survey. This work analyzed images obtained by the Canada-France-Hawaii Telescope (CFHT) and we are grateful for all the staff who worked to make this material possible. Finally, we thank the anonymous referee for very constructive suggestions and care in analyzing our work.

\section*{Data Availability}

All the catalogues and scripts used in this work may be required sending an email to marcelo.vicentin@usp.br. If there is any doubt, I am also available for contact and clarification. 



\bibliographystyle{mnras}
\bibliography{bib}

\newcommand{\noop}[1]{}
\begin{thebibliography}{}
\makeatletter
\relax
\def\mn@urlcharsother{\let\do\@makeother \do\$\do\&\do\#\do\^\do\_\do\%\do\~}
\def\mn@doi{\begingroup\mn@urlcharsother \@ifnextchar [ {\mn@doi@}
  {\mn@doi@[]}}
\def\mn@doi@[#1]#2{\def\@tempa{#1}\ifx\@tempa\@empty \href
  {http://dx.doi.org/#2} {doi:#2}\else \href {http://dx.doi.org/#2} {#1}\fi
  \endgroup}
\def\mn@eprint#1#2{\mn@eprint@#1:#2::\@nil}
\def\mn@eprint@arXiv#1{\href {http://arxiv.org/abs/#1} {{\tt arXiv:#1}}}
\def\mn@eprint@dblp#1{\href {http://dblp.uni-trier.de/rec/bibtex/#1.xml}
  {dblp:#1}}
\def\mn@eprint@#1:#2:#3:#4\@nil{\def\@tempa {#1}\def\@tempb {#2}\def\@tempc
  {#3}\ifx \@tempc \@empty \let \@tempc \@tempb \let \@tempb \@tempa \fi \ifx
  \@tempb \@empty \def\@tempb {arXiv}\fi \@ifundefined
  {mn@eprint@\@tempb}{\@tempb:\@tempc}{\expandafter \expandafter \csname
  mn@eprint@\@tempb\endcsname \expandafter{\@tempc}}}

\bibitem[\protect\citeauthoryear{{Adami} et~al.,}{{Adami}
  et~al.}{2010}]{adami10}
{Adami} C.,  et~al., 2010, \mn@doi [\aap] {10.1051/0004-6361/200913067}, \href
  {https://ui.adsabs.harvard.edu/abs/2010A&A...509A..81A} {509, A81}

\bibitem[\protect\citeauthoryear{{Adami} et~al.,}{{Adami}
  et~al.}{2011}]{adami11}
{Adami} C.,  et~al., 2011, \mn@doi [\aap] {10.1051/0004-6361/201015182}, \href
  {https://ui.adsabs.harvard.edu/abs/2011A&A...526A..18A} {526, A18}

\bibitem[\protect\citeauthoryear{{Aihara} et~al.,}{{Aihara}
  et~al.}{2018}]{aihara17}
{Aihara} H.,  et~al., 2018, \mn@doi [\pasj] {10.1093/pasj/psx066}, \href
  {https://ui.adsabs.harvard.edu/abs/2018PASJ...70S...4A} {70, S4}

\bibitem[\protect\citeauthoryear{{Almosallam}, {Jarvis}  \&
  {Roberts}}{{Almosallam} et~al.}{2016}]{almosallam16}
{Almosallam} I.~A.,  {Jarvis} M.~J.,   {Roberts} S.~J.,  2016, \mn@doi [\mnras]
  {10.1093/mnras/stw1618}, \href
  {https://ui.adsabs.harvard.edu/abs/2016MNRAS.462..726A} {462, 726}

\bibitem[\protect\citeauthoryear{{Araya-Araya}, {Vicentin}, {Sodr{\'e}},
  {Overzier}  \& {Cuevas}}{{Araya-Araya} et~al.}{2020}]{araya20}
{Araya-Araya} P.,  {Vicentin} M.~C.,  {Sodr{\'e}} L. J.,  {Overzier} R.~A.,
  {Cuevas} H.,  2020, {Protocluster detection in simulations of HSC-SSP and the
  10-year LSST forecast, using PCcones}, "submitted for publication"

\bibitem[\protect\citeauthoryear{{Arnouts} et~al.,}{{Arnouts}
  et~al.}{2002}]{arnouts02}
{Arnouts} S.,  et~al., 2002, \mn@doi [\mnras]
  {10.1046/j.1365-8711.2002.04988.x}, \href
  {https://ui.adsabs.harvard.edu/abs/2002MNRAS.329..355A} {329, 355}

\bibitem[\protect\citeauthoryear{{Bekki}}{{Bekki}}{1998}]{bekki98}
{Bekki} K.,  1998, \mn@doi [\apjl] {10.1086/311508}, \href
  {https://ui.adsabs.harvard.edu/abs/1998ApJ...502L.133B} {502, L133}

\bibitem[\protect\citeauthoryear{{Ben{\'\i}tez}}{{Ben{\'\i}tez}}{2000}]{benitez00}
{Ben{\'\i}tez} N.,  2000, \mn@doi [\apj] {10.1086/308947}, \href
  {https://ui.adsabs.harvard.edu/abs/2000ApJ...536..571B} {536, 571}

\bibitem[\protect\citeauthoryear{{Bertin}}{{Bertin}}{2006}]{bertin06}
{Bertin} E.,  2006, in {Gabriel} C.,  {Arviset} C.,  {Ponz} D.,   {Enrique} S.,
   eds,  Astronomical Society of the Pacific Conference Series Vol. 351,
  Astronomical Data Analysis Software and Systems XV. p.~112

\bibitem[\protect\citeauthoryear{{Bertin}}{{Bertin}}{2010}]{bertin10}
{Bertin} E.,  2010, {SWarp: Resampling and Co-adding FITS Images Together}
  (\mn@eprint {ascl} {1010.068})

\bibitem[\protect\citeauthoryear{{Bertin}}{{Bertin}}{2011}]{bertin2}
{Bertin} E.,  2011, in {Evans} I.~N.,  {Accomazzi} A.,  {Mink} D.~J.,   {Rots}
  A.~H.,  eds,  Astronomical Society of the Pacific Conference Series Vol. 442,
  Astronomical Data Analysis Software and Systems XX. p.~435

\bibitem[\protect\citeauthoryear{{Bertin} \& {Arnouts}}{{Bertin} \&
  {Arnouts}}{1996}]{bertin96}
{Bertin} E.,  {Arnouts} S.,  1996, \mn@doi [\aaps] {10.1051/aas:1996164}, \href
  {https://ui.adsabs.harvard.edu/abs/1996A&AS..117..393B} {117, 393}

\bibitem[\protect\citeauthoryear{{Bleem} et~al.,}{{Bleem}
  et~al.}{2015}]{bleem15}
{Bleem} L.~E.,  et~al., 2015, \mn@doi [\apjs] {10.1088/0067-0049/216/2/27},
  \href {https://ui.adsabs.harvard.edu/abs/2015ApJS..216...27B} {216, 27}

\bibitem[\protect\citeauthoryear{{Boris}, {Sodr{\'e}}, {Cypriano}, {Santos},
  {de Oliveira}  \& {West}}{{Boris} et~al.}{2007}]{boris}
{Boris} N.~V.,  {Sodr{\'e}} L. J.,  {Cypriano} E.~S.,  {Santos} W.~A.,  {de
  Oliveira} C.~M.,   {West} M.,  2007, \mn@doi [\apj] {10.1086/519992}, \href
  {https://ui.adsabs.harvard.edu/abs/2007ApJ...666..747B} {666, 747}

\bibitem[\protect\citeauthoryear{{Bower}, {Lucey}  \& {Ellis}}{{Bower}
  et~al.}{1992}]{bower92}
{Bower} R.~G.,  {Lucey} J.~R.,   {Ellis} R.~S.,  1992, \mn@doi [\mnras]
  {10.1093/mnras/254.4.601}, \href
  {https://ui.adsabs.harvard.edu/abs/1992MNRAS.254..601B} {254, 601}

\bibitem[\protect\citeauthoryear{{Brun} \& {Rademakers}}{{Brun} \&
  {Rademakers}}{1997}]{brun97}
{Brun} R.,  {Rademakers} F.,  1997, \mn@doi [Nuclear Instruments and Methods in
  Physics Research A] {10.1016/S0168-9002(97)00048-X}, \href
  {https://ui.adsabs.harvard.edu/abs/1997NIMPA.389...81B} {389, 81}

\bibitem[\protect\citeauthoryear{{Bruzual} \& {Charlot}}{{Bruzual} \&
  {Charlot}}{2003}]{bruzual03}
{Bruzual} G.,  {Charlot} S.,  2003, \mn@doi [\mnras]
  {10.1046/j.1365-8711.2003.06897.x}, \href
  {https://ui.adsabs.harvard.edu/abs/2003MNRAS.344.1000B} {344, 1000}

\bibitem[\protect\citeauthoryear{{B{\u{a}}descu}, {Yang}, {Bertoldi},
  {Zabludoff}, {Karim}  \& {Magnelli}}{{B{\u{a}}descu}
  et~al.}{2017}]{badescu17}
{B{\u{a}}descu} T.,  {Yang} Y.,  {Bertoldi} F.,  {Zabludoff} A.,  {Karim} A.,
  {Magnelli} B.,  2017, \mn@doi [\apj] {10.3847/1538-4357/aa8220}, \href
  {https://ui.adsabs.harvard.edu/abs/2017ApJ...845..172B} {845, 172}

\bibitem[\protect\citeauthoryear{{Capak} et~al.,}{{Capak}
  et~al.}{2011}]{capak11}
{Capak} P.~L.,  et~al., 2011, \mn@doi [\nat] {10.1038/nature09681}, \href
  {https://ui.adsabs.harvard.edu/abs/2011Natur.470..233C} {470, 233}

\bibitem[\protect\citeauthoryear{{Cardelli}, {Clayton}  \& {Mathis}}{{Cardelli}
  et~al.}{1989}]{cardelli89}
{Cardelli} J.~A.,  {Clayton} G.~C.,   {Mathis} J.~S.,  1989, \mn@doi [\apj]
  {10.1086/167900}, \href
  {https://ui.adsabs.harvard.edu/abs/1989ApJ...345..245C} {345, 245}

\bibitem[\protect\citeauthoryear{Charlor \& Bruzual}{Charlor \&
  Bruzual}{2007}]{charlot07}
Charlor S.,  Bruzual G.,  2007, in press

\bibitem[\protect\citeauthoryear{{Cheng} et~al.,}{{Cheng}
  et~al.}{2020}]{cheng20}
{Cheng} T.,  et~al., 2020, \mn@doi [\mnras] {10.1093/mnras/staa1096}, \href
  {https://ui.adsabs.harvard.edu/abs/2020MNRAS.494.5985C} {494, 5985}

\bibitem[\protect\citeauthoryear{{Chiang}, {Overzier}  \& {Gebhardt}}{{Chiang}
  et~al.}{2013}]{chiang13}
{Chiang} Y.-K.,  {Overzier} R.,   {Gebhardt} K.,  2013, \mn@doi [\apj]
  {10.1088/0004-637X/779/2/127}, \href
  {https://ui.adsabs.harvard.edu/abs/2013ApJ...779..127C} {779, 127}

\bibitem[\protect\citeauthoryear{{Chiang}, {Overzier}  \& {Gebhardt}}{{Chiang}
  et~al.}{2014}]{chiang14}
{Chiang} Y.-K.,  {Overzier} R.,   {Gebhardt} K.,  2014, \mn@doi [\apjl]
  {10.1088/2041-8205/782/1/L3}, \href
  {https://ui.adsabs.harvard.edu/abs/2014ApJ...782L...3C} {782, L3}

\bibitem[\protect\citeauthoryear{{Chiang} et~al.,}{{Chiang}
  et~al.}{2015}]{chiang15}
{Chiang} Y.-K.,  et~al., 2015, \mn@doi [\apj] {10.1088/0004-637X/808/1/37},
  \href {https://ui.adsabs.harvard.edu/abs/2015ApJ...808...37C} {808, 37}

\bibitem[\protect\citeauthoryear{{Coleman}, {Wu}  \& {Weedman}}{{Coleman}
  et~al.}{1980}]{coleman80}
{Coleman} G.~D.,  {Wu} C.~C.,   {Weedman} D.~W.,  1980, \mn@doi [\apjs]
  {10.1086/190674}, \href
  {https://ui.adsabs.harvard.edu/abs/1980ApJS...43..393C} {43, 393}

\bibitem[\protect\citeauthoryear{{Collister} \& {Lahav}}{{Collister} \&
  {Lahav}}{2004}]{collister04}
{Collister} A.~A.,  {Lahav} O.,  2004, \mn@doi [\pasp] {10.1086/383254}, \href
  {https://ui.adsabs.harvard.edu/abs/2004PASP..116..345C} {116, 345}

\bibitem[\protect\citeauthoryear{{Cooke}, {Hatch}, {Muldrew}, {Rigby}  \&
  {Kurk}}{{Cooke} et~al.}{2014}]{cooke14}
{Cooke} E.~A.,  {Hatch} N.~A.,  {Muldrew} S.~I.,  {Rigby} E.~E.,   {Kurk}
  J.~D.,  2014, \mn@doi [\mnras] {10.1093/mnras/stu522}, \href
  {https://ui.adsabs.harvard.edu/abs/2014MNRAS.440.3262C} {440, 3262}

\bibitem[\protect\citeauthoryear{{Croom} et~al.,}{{Croom}
  et~al.}{2009}]{croom09}
{Croom} S.~M.,  et~al., 2009, \mn@doi [\mnras]
  {10.1111/j.1365-2966.2008.14052.x}, \href
  {https://ui.adsabs.harvard.edu/abs/2009MNRAS.392...19C} {392, 19}

\bibitem[\protect\citeauthoryear{{Daddi} et~al.,}{{Daddi}
  et~al.}{2009}]{daddi09}
{Daddi} E.,  et~al., 2009, \mn@doi [\apj] {10.1088/0004-637X/694/2/1517}, \href
  {https://ui.adsabs.harvard.edu/abs/2009ApJ...694.1517D} {694, 1517}

\bibitem[\protect\citeauthoryear{{Danese}, {de Zotti}  \& {di Tullio}}{{Danese}
  et~al.}{1980}]{danese80}
{Danese} L.,  {de Zotti} G.,   {di Tullio} G.,  1980, \aap, \href
  {https://ui.adsabs.harvard.edu/abs/1980A&A....82..322D} {82, 322}

\bibitem[\protect\citeauthoryear{{Dannerbauer} et~al.,}{{Dannerbauer}
  et~al.}{2014}]{dannerbauer14}
{Dannerbauer} H.,  et~al., 2014, \mn@doi [\aap] {10.1051/0004-6361/201423771},
  \href {https://ui.adsabs.harvard.edu/abs/2014A&A...570A..55D} {570, A55}

\bibitem[\protect\citeauthoryear{{Dressler}}{{Dressler}}{1980}]{dressler80}
{Dressler} A.,  1980, \mn@doi [\apj] {10.1086/157753}, \href
  {https://ui.adsabs.harvard.edu/abs/1980ApJ...236..351D} {236, 351}

\bibitem[\protect\citeauthoryear{{Eftekharzadeh}, {Myers}, {Hennawi},
  {Djorgovski}, {Richards}, {Mahabal}  \& {Graham}}{{Eftekharzadeh}
  et~al.}{2017}]{eftekharzadeh17}
{Eftekharzadeh} S.,  {Myers} A.~D.,  {Hennawi} J.~F.,  {Djorgovski} S.~G.,
  {Richards} G.~T.,  {Mahabal} A.~A.,   {Graham} M.~J.,  2017, \mn@doi [\mnras]
  {10.1093/mnras/stx412}, \href
  {https://ui.adsabs.harvard.edu/abs/2017MNRAS.468...77E} {468, 77}

\bibitem[\protect\citeauthoryear{{Erben} et~al.,}{{Erben}
  et~al.}{2005}]{erben05}
{Erben} T.,  et~al., 2005, \mn@doi [Astronomische Nachrichten]
  {10.1002/asna.200510396}, \href
  {https://ui.adsabs.harvard.edu/abs/2005AN....326..432E} {326, 432}

\bibitem[\protect\citeauthoryear{{Farina}, {Falomo}  \& {Treves}}{{Farina}
  et~al.}{2011}]{farina11}
{Farina} E.~P.,  {Falomo} R.,   {Treves} A.,  2011, \mn@doi [\mnras]
  {10.1111/j.1365-2966.2011.18931.x}, \href
  {https://ui.adsabs.harvard.edu/abs/2011MNRAS.415.3163F} {415, 3163}

\bibitem[\protect\citeauthoryear{{Finkbeiner} et~al.,}{{Finkbeiner}
  et~al.}{2004}]{finkbeiner04}
{Finkbeiner} D.~P.,  et~al., 2004, \mn@doi [\aj] {10.1086/425050}, \href
  {https://ui.adsabs.harvard.edu/abs/2004AJ....128.2577F} {128, 2577}

\bibitem[\protect\citeauthoryear{{Fukugita}, {Ichikawa}, {Gunn}, {Doi},
  {Shimasaku}  \& {Schneider}}{{Fukugita} et~al.}{1996}]{fukugita96}
{Fukugita} M.,  {Ichikawa} T.,  {Gunn} J.~E.,  {Doi} M.,  {Shimasaku} K.,
  {Schneider} D.~P.,  1996, \mn@doi [\aj] {10.1086/117915}, \href
  {https://ui.adsabs.harvard.edu/abs/1996AJ....111.1748F} {111, 1748}

\bibitem[\protect\citeauthoryear{{Geach} et~al.,}{{Geach}
  et~al.}{2017}]{scuba17}
{Geach} J.~E.,  et~al., 2017, \mn@doi [\mnras] {10.1093/mnras/stw2721}, \href
  {https://ui.adsabs.harvard.edu/abs/2017MNRAS.465.1789G} {465, 1789}

\bibitem[\protect\citeauthoryear{{George} et~al.,}{{George}
  et~al.}{2011}]{george14}
{George} M.~R.,  et~al., 2011, \mn@doi [\apj] {10.1088/0004-637X/742/2/125},
  \href {https://ui.adsabs.harvard.edu/abs/2011ApJ...742..125G} {742, 125}

\bibitem[\protect\citeauthoryear{{Gonzalez} et~al.,}{{Gonzalez}
  et~al.}{2019}]{gonzalez19}
{Gonzalez} A.~H.,  et~al., 2019, \mn@doi [\apjs] {10.3847/1538-4365/aafad2},
  \href {https://ui.adsabs.harvard.edu/abs/2019ApJS..240...33G} {240, 33}

\bibitem[\protect\citeauthoryear{{Green}, {Myers}, {Barkhouse}, {Aldcroft},
  {Trichas}, {Richards}, {Ruiz}  \& {Hopkins}}{{Green} et~al.}{2011}]{green11}
{Green} P.~J.,  {Myers} A.~D.,  {Barkhouse} W.~A.,  {Aldcroft} T.~L.,
  {Trichas} M.,  {Richards} G.~T.,  {Ruiz} {\'A}.,   {Hopkins} P.~F.,  2011,
  \mn@doi [\apj] {10.1088/0004-637X/743/1/81}, \href
  {https://ui.adsabs.harvard.edu/abs/2011ApJ...743...81G} {743, 81}

\bibitem[\protect\citeauthoryear{{Gu}, {Fang}, {Yuan}  \& {Lu}}{{Gu}
  et~al.}{2020}]{gu20}
{Gu} Y.,  {Fang} G.,  {Yuan} Q.,   {Lu} S.,  2020, \mn@doi [\pasp]
  {10.1088/1538-3873/ab797d}, \href
  {https://ui.adsabs.harvard.edu/abs/2020PASP..132e4101G} {132, 054101}

\bibitem[\protect\citeauthoryear{{Gunn} \& {Gott}}{{Gunn} \&
  {Gott}}{1972}]{gunn72}
{Gunn} J.~E.,  {Gott} J.~Richard I.,  1972, \mn@doi [\apj] {10.1086/151605},
  \href {https://ui.adsabs.harvard.edu/abs/1972ApJ...176....1G} {176, 1}

\bibitem[\protect\citeauthoryear{{Hatch} et~al.,}{{Hatch}
  et~al.}{2011a}]{hatch11b}
{Hatch} N.~A.,  et~al., 2011a, \mn@doi [\mnras]
  {10.1111/j.1365-2966.2010.17538.x}, \href
  {https://ui.adsabs.harvard.edu/abs/2011MNRAS.410.1537H} {410, 1537}

\bibitem[\protect\citeauthoryear{{Hatch}, {Kurk}, {Pentericci}, {Venemans},
  {Kuiper}, {Miley}  \& {R{\"o}ttgering}}{{Hatch} et~al.}{2011b}]{hatch11a}
{Hatch} N.~A.,  {Kurk} J.~D.,  {Pentericci} L.,  {Venemans} B.~P.,  {Kuiper}
  E.,  {Miley} G.~K.,   {R{\"o}ttgering} H.~J.~A.,  2011b, \mn@doi [\mnras]
  {10.1111/j.1365-2966.2011.18735.x}, \href
  {https://ui.adsabs.harvard.edu/abs/2011MNRAS.415.2993H} {415, 2993}

\bibitem[\protect\citeauthoryear{{Hatch} et~al.,}{{Hatch} et~al.}{2014}]{hatch}
{Hatch} N.~A.,  et~al., 2014, \mn@doi [\mnras] {10.1093/mnras/stu1725}, \href
  {https://ui.adsabs.harvard.edu/abs/2014MNRAS.445..280H} {445, 280}

\bibitem[\protect\citeauthoryear{{Hayashi}, {Kodama}, {Tadaki}, {Koyama}  \&
  {Tanaka}}{{Hayashi} et~al.}{2012}]{hayashi12}
{Hayashi} M.,  {Kodama} T.,  {Tadaki} K.-i.,  {Koyama} Y.,   {Tanaka} I.,
  2012, \mn@doi [\apj] {10.1088/0004-637X/757/1/15}, \href
  {https://ui.adsabs.harvard.edu/abs/2012ApJ...757...15H} {757, 15}

\bibitem[\protect\citeauthoryear{{Henriques}, {White}, {Thomas}, {Angulo},
  {Guo}, {Lemson}, {Springel}  \& {Overzier}}{{Henriques}
  et~al.}{2015}]{henriques15}
{Henriques} B. M.~B.,  {White} S. D.~M.,  {Thomas} P.~A.,  {Angulo} R.,  {Guo}
  Q.,  {Lemson} G.,  {Springel} V.,   {Overzier} R.,  2015, \mn@doi [\mnras]
  {10.1093/mnras/stv705}, \href
  {https://ui.adsabs.harvard.edu/abs/2015MNRAS.451.2663H} {451, 2663}

\bibitem[\protect\citeauthoryear{{Higuchi} et~al.,}{{Higuchi}
  et~al.}{2019}]{higuchi19}
{Higuchi} R.,  et~al., 2019, \mn@doi [\apj] {10.3847/1538-4357/ab2192}, \href
  {https://ui.adsabs.harvard.edu/abs/2019ApJ...879...28H} {879, 28}

\bibitem[\protect\citeauthoryear{{Hilton} et~al.,}{{Hilton}
  et~al.}{2018}]{hilton18}
{Hilton} M.,  et~al., 2018, \mn@doi [\apjs] {10.3847/1538-4365/aaa6cb}, \href
  {https://ui.adsabs.harvard.edu/abs/2018ApJS..235...20H} {235, 20}

\bibitem[\protect\citeauthoryear{{Hoecker} et~al.,}{{Hoecker}
  et~al.}{2007}]{hoecker07}
{Hoecker} A.,  et~al., 2007, arXiv e-prints, \href
  {https://ui.adsabs.harvard.edu/abs/2007physics...3039H} {p. physics/0703039}

\bibitem[\protect\citeauthoryear{{Huang} et~al.,}{{Huang}
  et~al.}{2020}]{huang20}
{Huang} N.,  et~al., 2020, \mn@doi [\aj] {10.3847/1538-3881/ab6a96}, \href
  {https://ui.adsabs.harvard.edu/abs/2020AJ....159..110H} {159, 110}

\bibitem[\protect\citeauthoryear{{Ilbert} et~al.,}{{Ilbert}
  et~al.}{2006}]{ilbert06}
{Ilbert} O.,  et~al., 2006, \mn@doi [\aap] {10.1051/0004-6361:20065138}, \href
  {https://ui.adsabs.harvard.edu/abs/2006A&A...457..841I} {457, 841}

\bibitem[\protect\citeauthoryear{{Ivezi{\'c}}, {Connelly}, {Vand erPlas}  \&
  {Gray}}{{Ivezi{\'c}} et~al.}{2014}]{ivezic14}
{Ivezi{\'c}} {\v{Z}}.,  {Connelly} A.~J.,  {Vand erPlas} J.~T.,   {Gray} A.,
  2014, {Statistics, Data Mining, and Machine Learning in Astronomy}

\bibitem[\protect\citeauthoryear{{Jester} et~al.,}{{Jester}
  et~al.}{2005}]{jester05}
{Jester} S.,  et~al., 2005, \mn@doi [\aj] {10.1086/432466}, \href
  {https://ui.adsabs.harvard.edu/abs/2005AJ....130..873J} {130, 873}

\bibitem[\protect\citeauthoryear{{Jian} et~al.,}{{Jian} et~al.}{2014}]{jian14}
{Jian} H.-Y.,  et~al., 2014, \mn@doi [\apj] {10.1088/0004-637X/788/2/109},
  \href {https://ui.adsabs.harvard.edu/abs/2014ApJ...788..109J} {788, 109}

\bibitem[\protect\citeauthoryear{{Jian} et~al.,}{{Jian}
  et~al.}{2020}]{jian2020}
{Jian} H.-Y.,  et~al., 2020, \mn@doi [\apj] {10.3847/1538-4357/ab86a8}, \href
  {https://ui.adsabs.harvard.edu/abs/2020ApJ...894..125J} {894, 125}

\bibitem[\protect\citeauthoryear{{Joshi}, {Pillepich}, {Nelson}, {Marinacci},
  {Springel}, {Rodriguez-Gomez}, {Vogelsberger}  \& {Hernquist}}{{Joshi}
  et~al.}{2020}]{joshi20}
{Joshi} G.~D.,  {Pillepich} A.,  {Nelson} D.,  {Marinacci} F.,  {Springel} V.,
  {Rodriguez-Gomez} V.,  {Vogelsberger} M.,   {Hernquist} L.,  2020, \mn@doi
  [\mnras] {10.1093/mnras/staa1668}, \href
  {https://ui.adsabs.harvard.edu/abs/2020MNRAS.tmp.1796J} {}

\bibitem[\protect\citeauthoryear{{Kaiser}}{{Kaiser}}{1984}]{kaiser84}
{Kaiser} N.,  1984, \mn@doi [\apjl] {10.1086/184341}, \href
  {https://ui.adsabs.harvard.edu/abs/1984ApJ...284L...9K} {284, L9}

\bibitem[\protect\citeauthoryear{{Kodama}, {Arimoto}, {Barger}  \&
  {Arag'on-Salamanca}}{{Kodama} et~al.}{1998}]{kodama98}
{Kodama} T.,  {Arimoto} N.,  {Barger} A.~J.,   {Arag'on-Salamanca} A.,  1998,
  \aap, \href {https://ui.adsabs.harvard.edu/abs/1998A&A...334...99K} {334, 99}

\bibitem[\protect\citeauthoryear{{Koutsouridou} \& {Cattaneo}}{{Koutsouridou}
  \& {Cattaneo}}{2019}]{kout19}
{Koutsouridou} I.,  {Cattaneo} A.,  2019, \mn@doi [\mnras]
  {10.1093/mnras/stz2916}, \href
  {https://ui.adsabs.harvard.edu/abs/2019MNRAS.490.5375K} {490, 5375}

\bibitem[\protect\citeauthoryear{{Kravtsov} \& {Borgani}}{{Kravtsov} \&
  {Borgani}}{2012}]{kravtsov12}
{Kravtsov} A.~V.,  {Borgani} S.,  2012, \mn@doi [\araa]
  {10.1146/annurev-astro-081811-125502}, \href
  {https://ui.adsabs.harvard.edu/abs/2012ARA&A..50..353K} {50, 353}

\bibitem[\protect\citeauthoryear{{Krefting} et~al.,}{{Krefting}
  et~al.}{2020}]{krefting20}
{Krefting} N.,  et~al., 2020, \mn@doi [\apj] {10.3847/1538-4357/ab60a0}, \href
  {https://ui.adsabs.harvard.edu/abs/2020ApJ...889..185K} {889, 185}

\bibitem[\protect\citeauthoryear{{Kuiper}, {Venemans}, {Hatch}, {Miley}  \&
  {R{\"o}ttgering}}{{Kuiper} et~al.}{2012}]{kuiper12}
{Kuiper} E.,  {Venemans} B.~P.,  {Hatch} N.~A.,  {Miley} G.~K.,
  {R{\"o}ttgering} H.~J.~A.,  2012, \mn@doi [\mnras]
  {10.1111/j.1365-2966.2012.20800.x}, \href
  {https://ui.adsabs.harvard.edu/abs/2012MNRAS.425..801K} {425, 801}

\bibitem[\protect\citeauthoryear{{Laigle} et~al.,}{{Laigle}
  et~al.}{2016}]{laigle16}
{Laigle} C.,  et~al., 2016, \mn@doi [\apjs] {10.3847/0067-0049/224/2/24}, \href
  {https://ui.adsabs.harvard.edu/abs/2016ApJS..224...24L} {224, 24}

\bibitem[\protect\citeauthoryear{{Lang}}{{Lang}}{2014}]{lang14}
{Lang} D.,  2014, \mn@doi [\aj] {10.1088/0004-6256/147/5/108}, \href
  {https://ui.adsabs.harvard.edu/abs/2014AJ....147..108L} {147, 108}

\bibitem[\protect\citeauthoryear{{Liu} et~al.,}{{Liu} et~al.}{2019}]{liu19}
{Liu} D.,  et~al., 2019, \mn@doi [\apj] {10.3847/1538-4357/ab578d}, \href
  {https://ui.adsabs.harvard.edu/abs/2019ApJ...887..235L} {887, 235}

\bibitem[\protect\citeauthoryear{{Lu}, {Cappellari}, {Mao}, {Ge}  \& {Li}}{{Lu}
  et~al.}{2020}]{lu20}
{Lu} S.,  {Cappellari} M.,  {Mao} S.,  {Ge} J.,   {Li} R.,  2020, \mn@doi
  [\mnras] {10.1093/mnras/staa1481}, \href
  {https://ui.adsabs.harvard.edu/abs/2020MNRAS.495.4820L} {495, 4820}

\bibitem[\protect\citeauthoryear{{Maraston}}{{Maraston}}{2005}]{maraston05}
{Maraston} C.,  2005, \mn@doi [\mnras] {10.1111/j.1365-2966.2005.09270.x},
  \href {https://ui.adsabs.harvard.edu/abs/2005MNRAS.362..799M} {362, 799}

\bibitem[\protect\citeauthoryear{{Martinache} et~al.,}{{Martinache}
  et~al.}{2018}]{martinache18}
{Martinache} C.,  et~al., 2018, \mn@doi [\aap] {10.1051/0004-6361/201833198},
  \href {https://ui.adsabs.harvard.edu/abs/2018A&A...620A.198M} {620, A198}

\bibitem[\protect\citeauthoryear{{Mehrtens} et~al.,}{{Mehrtens}
  et~al.}{2012}]{mehrtens12}
{Mehrtens} N.,  et~al., 2012, \mn@doi [\mnras]
  {10.1111/j.1365-2966.2012.20931.x}, \href
  {https://ui.adsabs.harvard.edu/abs/2012MNRAS.423.1024M} {423, 1024}

\bibitem[\protect\citeauthoryear{{Mei} et~al.,}{{Mei} et~al.}{2006}]{mei06}
{Mei} S.,  et~al., 2006, \mn@doi [\apj] {10.1086/499259}, \href
  {https://ui.adsabs.harvard.edu/abs/2006ApJ...639...81M} {639, 81}

\bibitem[\protect\citeauthoryear{{Meisner}, {Lang}  \& {Schlegel}}{{Meisner}
  et~al.}{2017a}]{meisner17a}
{Meisner} A.~M.,  {Lang} D.,   {Schlegel} D.~J.,  2017a, \mn@doi [\aj]
  {10.3847/1538-3881/153/1/38}, \href
  {https://ui.adsabs.harvard.edu/abs/2017AJ....153...38M} {153, 38}

\bibitem[\protect\citeauthoryear{{Meisner}, {Lang}  \& {Schlegel}}{{Meisner}
  et~al.}{2017b}]{meisner17b}
{Meisner} A.~M.,  {Lang} D.,   {Schlegel} D.~J.,  2017b, \mn@doi [\aj]
  {10.3847/1538-3881/aa894e}, \href
  {https://ui.adsabs.harvard.edu/abs/2017AJ....154..161M} {154, 161}

\bibitem[\protect\citeauthoryear{{Meisner}, {Lang}  \& {Schlegel}}{{Meisner}
  et~al.}{2018}]{meisner18a}
{Meisner} A.~M.,  {Lang} D.,   {Schlegel} D.~J.,  2018, \mn@doi [\aj]
  {10.3847/1538-3881/aacbcd}, \href
  {https://ui.adsabs.harvard.edu/abs/2018AJ....156...69M} {156, 69}

\bibitem[\protect\citeauthoryear{{Molino} et~al.,}{{Molino}
  et~al.}{2017}]{molino17}
{Molino} A.,  et~al., 2017, \mn@doi [\mnras] {10.1093/mnras/stx1243}, \href
  {https://ui.adsabs.harvard.edu/abs/2017MNRAS.470...95M} {470, 95}

\bibitem[\protect\citeauthoryear{{Moore}, {Katz}, {Lake}, {Dressler}  \&
  {Oemler}}{{Moore} et~al.}{1996}]{moore96}
{Moore} B.,  {Katz} N.,  {Lake} G.,  {Dressler} A.,   {Oemler} A.,  1996,
  \mn@doi [\nat] {10.1038/379613a0}, \href
  {https://ui.adsabs.harvard.edu/abs/1996Natur.379..613M} {379, 613}

\bibitem[\protect\citeauthoryear{{Mullis}, {Rosati}, {Lamer}, {B{\"o}hringer},
  {Schwope}, {Schuecker}  \& {Fassbender}}{{Mullis} et~al.}{2005}]{mullis05}
{Mullis} C.~R.,  {Rosati} P.,  {Lamer} G.,  {B{\"o}hringer} H.,  {Schwope} A.,
  {Schuecker} P.,   {Fassbender} R.,  2005, \mn@doi [\apjl] {10.1086/429801},
  \href {https://ui.adsabs.harvard.edu/abs/2005ApJ...623L..85M} {623, L85}

\bibitem[\protect\citeauthoryear{{Noirot} et~al.,}{{Noirot}
  et~al.}{2017}]{carla_survey}
{Noirot} G.,  et~al., 2017, in Galaxy Evolution Across Time. p.~72,
  \mn@doi{10.5281/zenodo.809356}

\bibitem[\protect\citeauthoryear{{Onoue} et~al.,}{{Onoue}
  et~al.}{2018}]{onoue18}
{Onoue} M.,  et~al., 2018, \mn@doi [\pasj] {10.1093/pasj/psx092}, \href
  {https://ui.adsabs.harvard.edu/abs/2018PASJ...70S..31O} {70, S31}

\bibitem[\protect\citeauthoryear{{Overzier}}{{Overzier}}{2016}]{overzier16}
{Overzier} R.~A.,  2016, \mn@doi [\aapr] {10.1007/s00159-016-0100-3}, \href
  {https://ui.adsabs.harvard.edu/abs/2016A&ARv..24...14O} {24, 14}

\bibitem[\protect\citeauthoryear{{Overzier} et~al.,}{{Overzier}
  et~al.}{2006}]{overzier06}
{Overzier} R.~A.,  et~al., 2006, \mn@doi [\apj] {10.1086/498234}, \href
  {https://ui.adsabs.harvard.edu/abs/2006ApJ...637...58O} {637, 58}

\bibitem[\protect\citeauthoryear{{Overzier} et~al.,}{{Overzier}
  et~al.}{2008}]{overzier08}
{Overzier} R.~A.,  et~al., 2008, \mn@doi [\apj] {10.1086/524342}, \href
  {https://ui.adsabs.harvard.edu/abs/2008ApJ...673..143O} {673, 143}

\bibitem[\protect\citeauthoryear{{Oyaizu}, {Lima}, {Cunha}, {Lin}  \&
  {Frieman}}{{Oyaizu} et~al.}{2008}]{oyaizu08}
{Oyaizu} H.,  {Lima} M.,  {Cunha} C.~E.,  {Lin} H.,   {Frieman} J.,  2008,
  \mn@doi [\apj] {10.1086/592591}, \href
  {https://ui.adsabs.harvard.edu/abs/2008ApJ...689..709O} {689, 709}

\bibitem[\protect\citeauthoryear{{Pier}, {Munn}, {Hindsley}, {Hennessy},
  {Kent}, {Lupton}  \& {Ivezi{\'c}}}{{Pier} et~al.}{2003}]{pier03}
{Pier} J.~R.,  {Munn} J.~A.,  {Hindsley} R.~B.,  {Hennessy} G.~S.,  {Kent}
  S.~M.,  {Lupton} R.~H.,   {Ivezi{\'c}} {\v{Z}}.,  2003, \mn@doi [\aj]
  {10.1086/346138}, \href
  {https://ui.adsabs.harvard.edu/abs/2003AJ....125.1559P} {125, 1559}

\bibitem[\protect\citeauthoryear{{Planck Collaboration} et~al.,}{{Planck
  Collaboration} et~al.}{2014}]{planck1}
{Planck Collaboration} et~al., 2014, \mn@doi [\aap]
  {10.1051/0004-6361/201321591}, \href
  {https://ui.adsabs.harvard.edu/abs/2014A&A...571A..16P} {571, A16}

\bibitem[\protect\citeauthoryear{{Planck Collaboration} et~al.,}{{Planck
  Collaboration} et~al.}{2015}]{planck15}
{Planck Collaboration} et~al., 2015, \mn@doi [\aap]
  {10.1051/0004-6361/201424790}, \href
  {https://ui.adsabs.harvard.edu/abs/2015A&A...582A..30P} {582, A30}

\bibitem[\protect\citeauthoryear{{Planck Collaboration} et~al.,}{{Planck
  Collaboration} et~al.}{2016}]{planck16}
{Planck Collaboration} et~al., 2016, \mn@doi [\aap]
  {10.1051/0004-6361/201527206}, \href
  {https://ui.adsabs.harvard.edu/abs/2016A&A...596A.100P} {596, A100}

\bibitem[\protect\citeauthoryear{{Rigby} et~al.,}{{Rigby}
  et~al.}{2014}]{rigby14}
{Rigby} E.~E.,  et~al., 2014, \mn@doi [\mnras] {10.1093/mnras/stt2019}, \href
  {https://ui.adsabs.harvard.edu/abs/2014MNRAS.437.1882R} {437, 1882}

\bibitem[\protect\citeauthoryear{{Rosati}, {Borgani}  \& {Norman}}{{Rosati}
  et~al.}{2002}]{rosati02}
{Rosati} P.,  {Borgani} S.,   {Norman} C.,  2002, \mn@doi [\araa]
  {10.1146/annurev.astro.40.120401.150547}, \href
  {https://ui.adsabs.harvard.edu/abs/2002ARA&A..40..539R} {40, 539}

\bibitem[\protect\citeauthoryear{{Sadeh}, {Abdalla}  \& {Lahav}}{{Sadeh}
  et~al.}{2016}]{sadeh16}
{Sadeh} I.,  {Abdalla} F.~B.,   {Lahav} O.,  2016, \mn@doi [\pasp]
  {10.1088/1538-3873/128/968/104502}, \href
  {https://ui.adsabs.harvard.edu/abs/2016PASP..128j4502S} {128, 104502}

\bibitem[\protect\citeauthoryear{{S{\'a}nchez} \&
  {Gonz{\'a}lez-Serrano}}{{S{\'a}nchez} \&
  {Gonz{\'a}lez-Serrano}}{1999}]{sanchez99}
{S{\'a}nchez} S.~F.,  {Gonz{\'a}lez-Serrano} J.~I.,  1999, \aap, \href
  {https://ui.adsabs.harvard.edu/abs/1999A&A...352..383S} {352, 383}

\bibitem[\protect\citeauthoryear{{S{\'a}nchez} \&
  {Gonz{\'a}lez-Serrano}}{{S{\'a}nchez} \&
  {Gonz{\'a}lez-Serrano}}{2002}]{sanchez02}
{S{\'a}nchez} S.~F.,  {Gonz{\'a}lez-Serrano} J.~I.,  2002, \mn@doi [\aap]
  {10.1051/0004-6361:20021455}, \href
  {https://ui.adsabs.harvard.edu/abs/2002A&A...396..773S} {396, 773}

\bibitem[\protect\citeauthoryear{{Sandrinelli}, {Falomo}, {Treves}, {Farina}
  \& {Uslenghi}}{{Sandrinelli} et~al.}{2014}]{sandrinelli14}
{Sandrinelli} A.,  {Falomo} R.,  {Treves} A.,  {Farina} E.~P.,   {Uslenghi} M.,
   2014, \mn@doi [\mnras] {10.1093/mnras/stu1526}, \href
  {https://ui.adsabs.harvard.edu/abs/2014MNRAS.444.1835S} {444, 1835}

\bibitem[\protect\citeauthoryear{{Schirmer}}{{Schirmer}}{2013}]{schirmer13}
{Schirmer} M.,  2013, \mn@doi [\apjs] {10.1088/0067-0049/209/2/21}, \href
  {https://ui.adsabs.harvard.edu/abs/2013ApJS..209...21S} {209, 21}

\bibitem[\protect\citeauthoryear{{Schlafly} et~al.,}{{Schlafly}
  et~al.}{2018}]{schlafly18}
{Schlafly} E.~F.,  et~al., 2018, \mn@doi [\apjs] {10.3847/1538-4365/aaa3e2},
  \href {https://ui.adsabs.harvard.edu/abs/2018ApJS..234...39S} {234, 39}

\bibitem[\protect\citeauthoryear{{Schlafly}, {Meisner}  \& {Green}}{{Schlafly}
  et~al.}{2019}]{schlafly19}
{Schlafly} E.~F.,  {Meisner} A.~M.,   {Green} G.~M.,  2019, \mn@doi [\apjs]
  {10.3847/1538-4365/aafbea}, \href
  {https://ui.adsabs.harvard.edu/abs/2019ApJS..240...30S} {240, 30}

\bibitem[\protect\citeauthoryear{{Schlegel}, {Finkbeiner}  \&
  {Davis}}{{Schlegel} et~al.}{1998}]{schlegel98}
{Schlegel} D.~J.,  {Finkbeiner} D.~P.,   {Davis} M.,  1998, \mn@doi [\apj]
  {10.1086/305772}, \href
  {https://ui.adsabs.harvard.edu/abs/1998ApJ...500..525S} {500, 525}

\bibitem[\protect\citeauthoryear{{Seymour}, {Stern}  \& {De Breuck}}{{Seymour}
  et~al.}{2007}]{seymour07}
{Seymour} N.,  {Stern} D.,   {De Breuck} C.,  2007, in {Afonso} J.,  {Ferguson}
  H.~C.,  {Mobasher} B.,   {Norris} R.,  eds,  Astronomical Society of the
  Pacific Conference Series Vol. 380, Deepest Astronomical Surveys. p.~393
  (\mn@eprint {arXiv} {astro-ph/0604226})

\bibitem[\protect\citeauthoryear{{Shen} et~al.,}{{Shen} et~al.}{2011}]{shen11}
{Shen} Y.,  et~al., 2011, \mn@doi [\apjs] {10.1088/0067-0049/194/2/45}, \href
  {https://ui.adsabs.harvard.edu/abs/2011ApJS..194...45S} {194, 45}

\bibitem[\protect\citeauthoryear{{Springel} et~al.,}{{Springel}
  et~al.}{2005}]{springel05}
{Springel} V.,  et~al., 2005, \mn@doi [\nat] {10.1038/nature03597}, \href
  {https://ui.adsabs.harvard.edu/abs/2005Natur.435..629S} {435, 629}

\bibitem[\protect\citeauthoryear{{Stanford} et~al.,}{{Stanford}
  et~al.}{2006}]{stanford06}
{Stanford} S.~A.,  et~al., 2006, \mn@doi [\apjl] {10.1086/506449}, \href
  {https://ui.adsabs.harvard.edu/abs/2006ApJ...646L..13S} {646, L13}

\bibitem[\protect\citeauthoryear{{Stott} et~al.,}{{Stott}
  et~al.}{2020}]{stott20}
{Stott} J.~P.,  et~al., 2020, arXiv e-prints, \href
  {https://ui.adsabs.harvard.edu/abs/2020arXiv200607384S} {p. arXiv:2006.07384}

\bibitem[\protect\citeauthoryear{{Tanaka}}{{Tanaka}}{2015}]{tanaka15}
{Tanaka} M.,  2015, \mn@doi [\apj] {10.1088/0004-637X/801/1/20}, \href
  {https://ui.adsabs.harvard.edu/abs/2015ApJ...801...20T} {801, 20}

\bibitem[\protect\citeauthoryear{{Taylor}}{{Taylor}}{2005}]{taylor05}
{Taylor} M.~B.,  2005, {TOPCAT \&amp; STIL: Starlink Table/VOTable Processing
  Software}.
p.~29

\bibitem[\protect\citeauthoryear{{Tiley} et~al.,}{{Tiley}
  et~al.}{2020}]{tiley20}
{Tiley} A.~L.,  et~al., 2020, \mn@doi [\mnras] {10.1093/mnras/staa1418}, \href
  {https://ui.adsabs.harvard.edu/abs/2020MNRAS.tmp.1548T} {}

\bibitem[\protect\citeauthoryear{{Toshikawa} et~al.,}{{Toshikawa}
  et~al.}{2018}]{toshikawa18}
{Toshikawa} J.,  et~al., 2018, \mn@doi [\pasj] {10.1093/pasj/psx102}, \href
  {https://ui.adsabs.harvard.edu/abs/2018PASJ...70S..12T} {70, S12}

\bibitem[\protect\citeauthoryear{{Trussler}, {Maiolino}, {Maraston}, {Peng},
  {Thomas}, {Goddard}  \& {Lian}}{{Trussler} et~al.}{2020}]{trussler20}
{Trussler} J.,  {Maiolino} R.,  {Maraston} C.,  {Peng} Y.,  {Thomas} D.,
  {Goddard} D.,   {Lian} J.,  2020, arXiv e-prints, \href
  {https://ui.adsabs.harvard.edu/abs/2020arXiv200601154T} {p. arXiv:2006.01154}

\bibitem[\protect\citeauthoryear{{Venemans} et~al.,}{{Venemans}
  et~al.}{2002}]{venemans02}
{Venemans} B.~P.,  et~al., 2002, \mn@doi [\apjl] {10.1086/340563}, \href
  {https://ui.adsabs.harvard.edu/abs/2002ApJ...569L..11V} {569, L11}

\bibitem[\protect\citeauthoryear{{Venemans} et~al.,}{{Venemans}
  et~al.}{2007}]{venemans07}
{Venemans} B.~P.,  et~al., 2007, \mn@doi [\aap] {10.1051/0004-6361:20053941},
  \href {https://ui.adsabs.harvard.edu/abs/2007A&A...461..823V} {461, 823}

\bibitem[\protect\citeauthoryear{{V{\'e}ron-Cetty} \&
  {V{\'e}ron}}{{V{\'e}ron-Cetty} \& {V{\'e}ron}}{2010}]{veron10}
{V{\'e}ron-Cetty} M.~P.,  {V{\'e}ron} P.,  2010, \mn@doi [\aap]
  {10.1051/0004-6361/201014188}, \href
  {https://ui.adsabs.harvard.edu/abs/2010A&A...518A..10V} {518, A10}

\bibitem[\protect\citeauthoryear{{Visvanathan} \& {Sandage}}{{Visvanathan} \&
  {Sandage}}{1977}]{visvanathan77}
{Visvanathan} N.,  {Sandage} A.,  1977, \mn@doi [\apj] {10.1086/155464}, \href
  {https://ui.adsabs.harvard.edu/abs/1977ApJ...216..214V} {216, 214}

\bibitem[\protect\citeauthoryear{{Wen} \& {Han}}{{Wen} \& {Han}}{2011}]{wen11}
{Wen} Z.~L.,  {Han} J.~L.,  2011, \mn@doi [\apj] {10.1088/0004-637X/734/1/68},
  \href {https://ui.adsabs.harvard.edu/abs/2011ApJ...734...68W} {734, 68}

\bibitem[\protect\citeauthoryear{{White} \& {Frenk}}{{White} \&
  {Frenk}}{1991}]{white91}
{White} S. D.~M.,  {Frenk} C.~S.,  1991, \mn@doi [\apj] {10.1086/170483}, \href
  {https://ui.adsabs.harvard.edu/abs/1991ApJ...379...52W} {379, 52}

\bibitem[\protect\citeauthoryear{{White} \& {Rees}}{{White} \&
  {Rees}}{1978}]{white78}
{White} S.~D.~M.,  {Rees} M.~J.,  1978, \mn@doi [\mnras]
  {10.1093/mnras/183.3.341}, \href
  {https://ui.adsabs.harvard.edu/abs/1978MNRAS.183..341W} {183, 341}

\bibitem[\protect\citeauthoryear{{Wylezalek} et~al.,}{{Wylezalek}
  et~al.}{2013}]{wylezalek13}
{Wylezalek} D.,  et~al., 2013, \mn@doi [\apj] {10.1088/0004-637X/769/1/79},
  \href {https://ui.adsabs.harvard.edu/abs/2013ApJ...769...79W} {769, 79}

\bibitem[\protect\citeauthoryear{{Zel'dovich} \& {Syunyaev}}{{Zel'dovich} \&
  {Syunyaev}}{1972}]{zeldovich72}
{Zel'dovich} Y.~B.,  {Syunyaev} R.~A.,  1972, Soviet Journal of Experimental
  and Theoretical Physics, \href
  {https://ui.adsabs.harvard.edu/abs/1972JETP...35...81Z} {35, 81}

\bibitem[\protect\citeauthoryear{{van Dokkum}}{{van
  Dokkum}}{2005}]{vandokkum05}
{van Dokkum} P.~G.,  2005, \mn@doi [\aj] {10.1086/497593}, \href
  {https://ui.adsabs.harvard.edu/abs/2005AJ....130.2647V} {130, 2647}

\bibitem[\protect\citeauthoryear{{van den Bosch}, {Aquino}, {Yang}, {Mo},
  {Pasquali}, {McIntosh}, {Weinmann}  \& {Kang}}{{van den Bosch}
  et~al.}{2008}]{vandenbosch08}
{van den Bosch} F.~C.,  {Aquino} D.,  {Yang} X.,  {Mo} H.~J.,  {Pasquali} A.,
  {McIntosh} D.~H.,  {Weinmann} S.~M.,   {Kang} X.,  2008, \mn@doi [\mnras]
  {10.1111/j.1365-2966.2008.13230.x}, 387, 79

\bibitem[\protect\citeauthoryear{{van der Burg} et~al.,}{{van der Burg}
  et~al.}{2020}]{vanderburg20}
{van der Burg} R. F.~J.,  et~al., 2020, arXiv e-prints, \href
  {https://ui.adsabs.harvard.edu/abs/2020arXiv200410757V} {p. arXiv:2004.10757}

\makeatother
\end{thebibliography}


\appendix
\section{QSOs available spectra}
\label{sec:apA}

\begin{figure*}
\centering
\includegraphics[width=\textwidth, height=10cm, trim= 0 0 0 0cm]{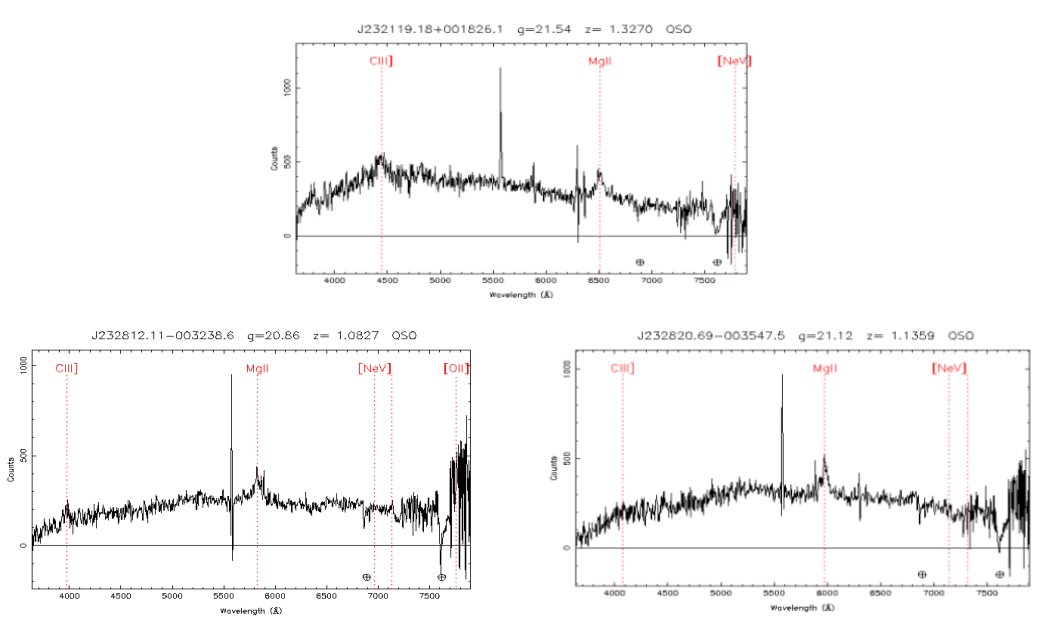}
\caption{\small Upper panel: Spectrum of the SDSSJ232119+001826.1 in the S5 field; Lower panel: Spectra of the SDSSJ232812-003238.6 and SDSSJ232821-003547.5, in the S6 field}.
\label{fig:apendice_spec4}
\end{figure*}

Here we present the spectra of our quasars sample (Figures \ref{fig:apendice_spec1}, \ref{fig:apendice_spec2}, \ref{fig:apendice_spec3}, and \ref{fig:apendice_spec4}).  The firt three figures show the spectra obtained by the SDSS and can be consulted at the website\footnote{\url{http://skyserver.sdss.org/dr15/en/tools/chart/list.aspx}}. There is one case in field S5, and two others in S6, which SDSS do not have this data. However, they were observed spectroscopically by \citet{croom09} and, although the network system for this survey is offline, Professor Scott Croom kindly shared the spectra images and they are presented at Figure \ref{fig:apendice_spec4}.

\begin{figure*}
\centering
\includegraphics[width=\textwidth, height=20cm, trim= 0 2cm 0 0cm]{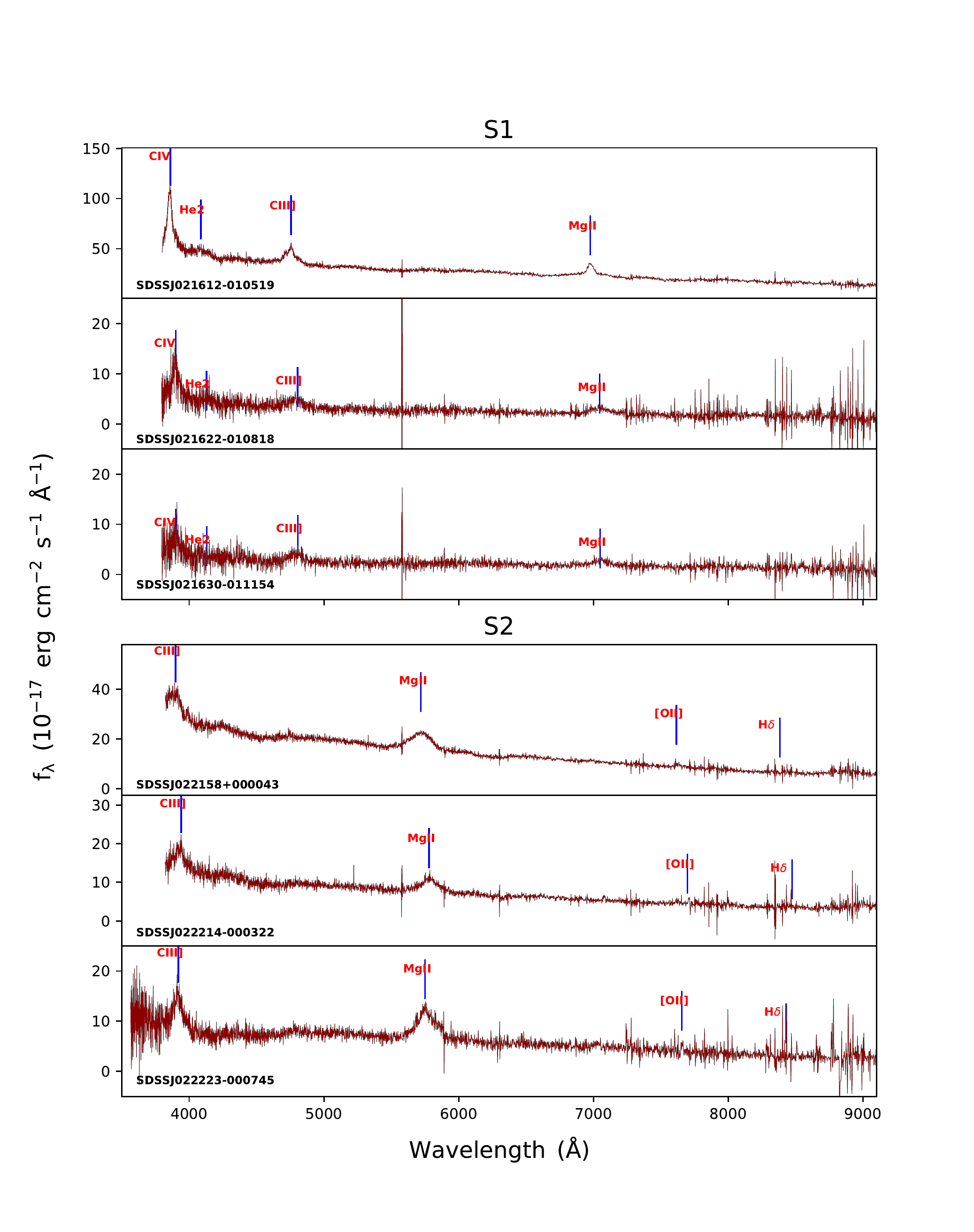}
\caption{\small Available spectra of the QSOs.}
\label{fig:apendice_spec1}
\end{figure*}

\begin{figure*}
\centering
\includegraphics[width=\textwidth, height=20cm, trim= 0 2cm 0 0cm]{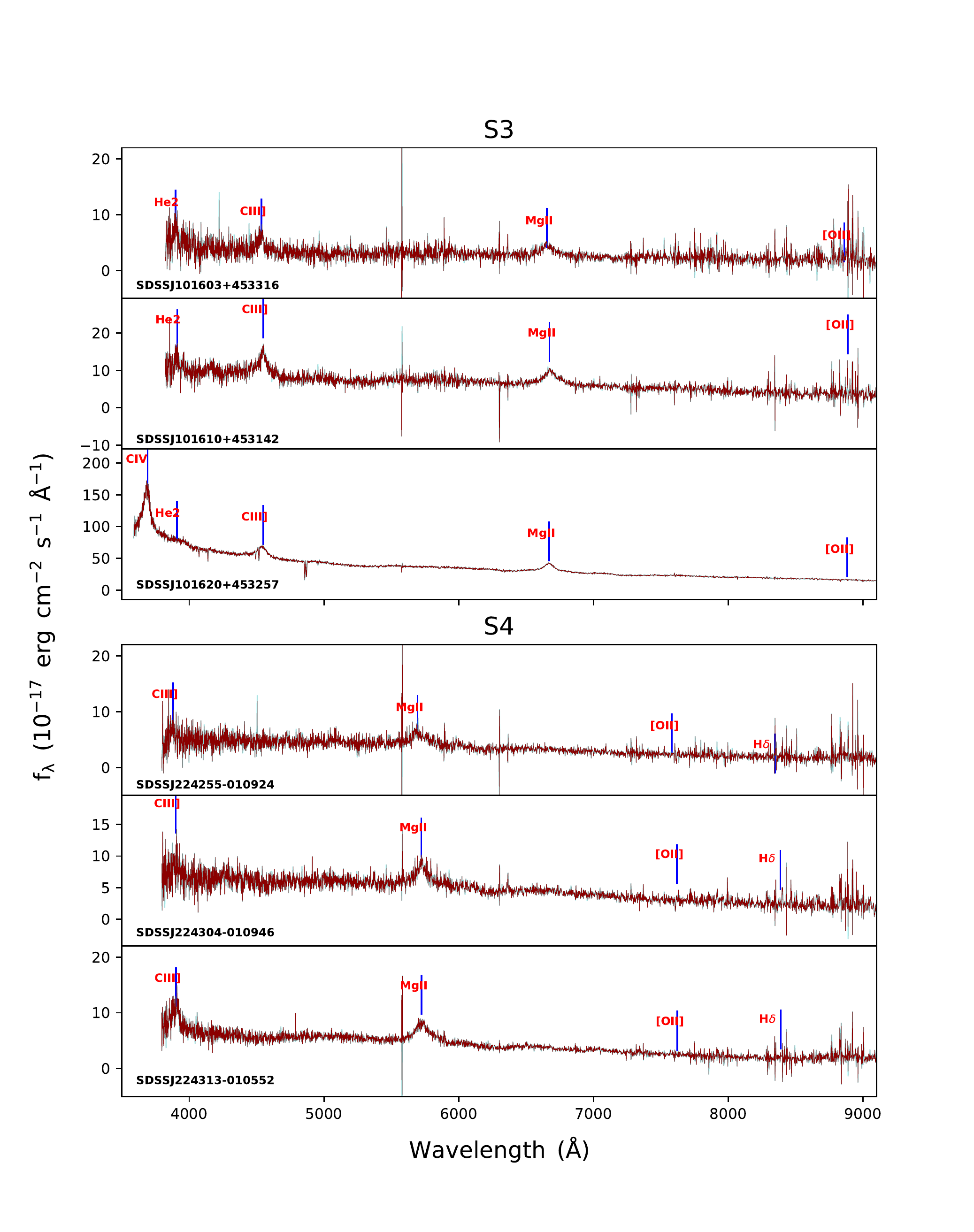}
\caption{\small Continuation of Figure \ref{fig:apendice_spec1}}.
\label{fig:apendice_spec2}
\end{figure*}

\begin{figure*}
\centering
\includegraphics[width=\textwidth, height=10cm, trim= 0 0 0 0cm]{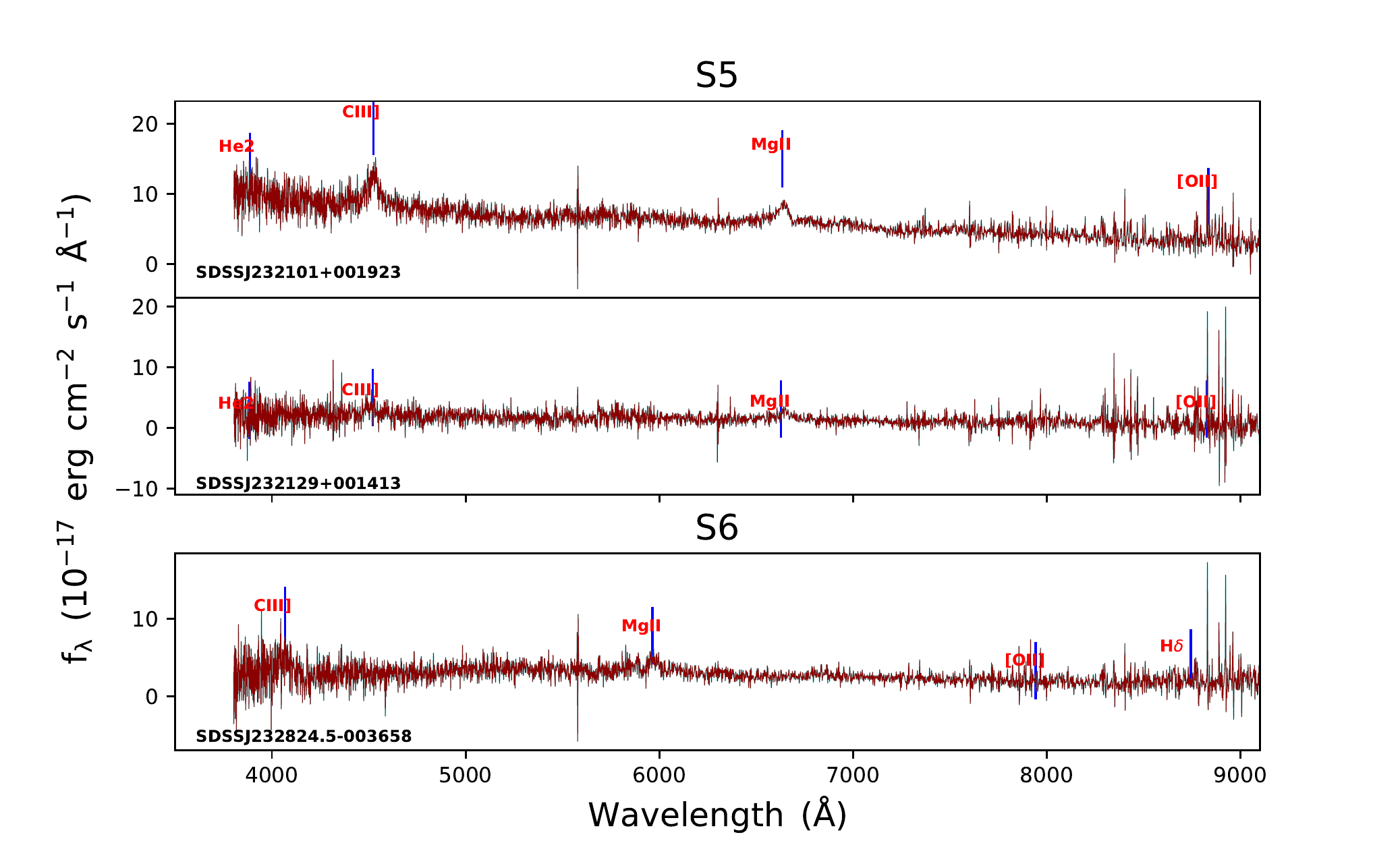}
\caption{\small Continuation of Figure \ref{fig:apendice_spec2}}.
\label{fig:apendice_spec3}
\end{figure*}

\section{Galaxy counts}
\label{sec:apB}

\begin{figure*}
\centering
\includegraphics[width=\textwidth, height=9.0in, trim= 0 2cm 0 0]{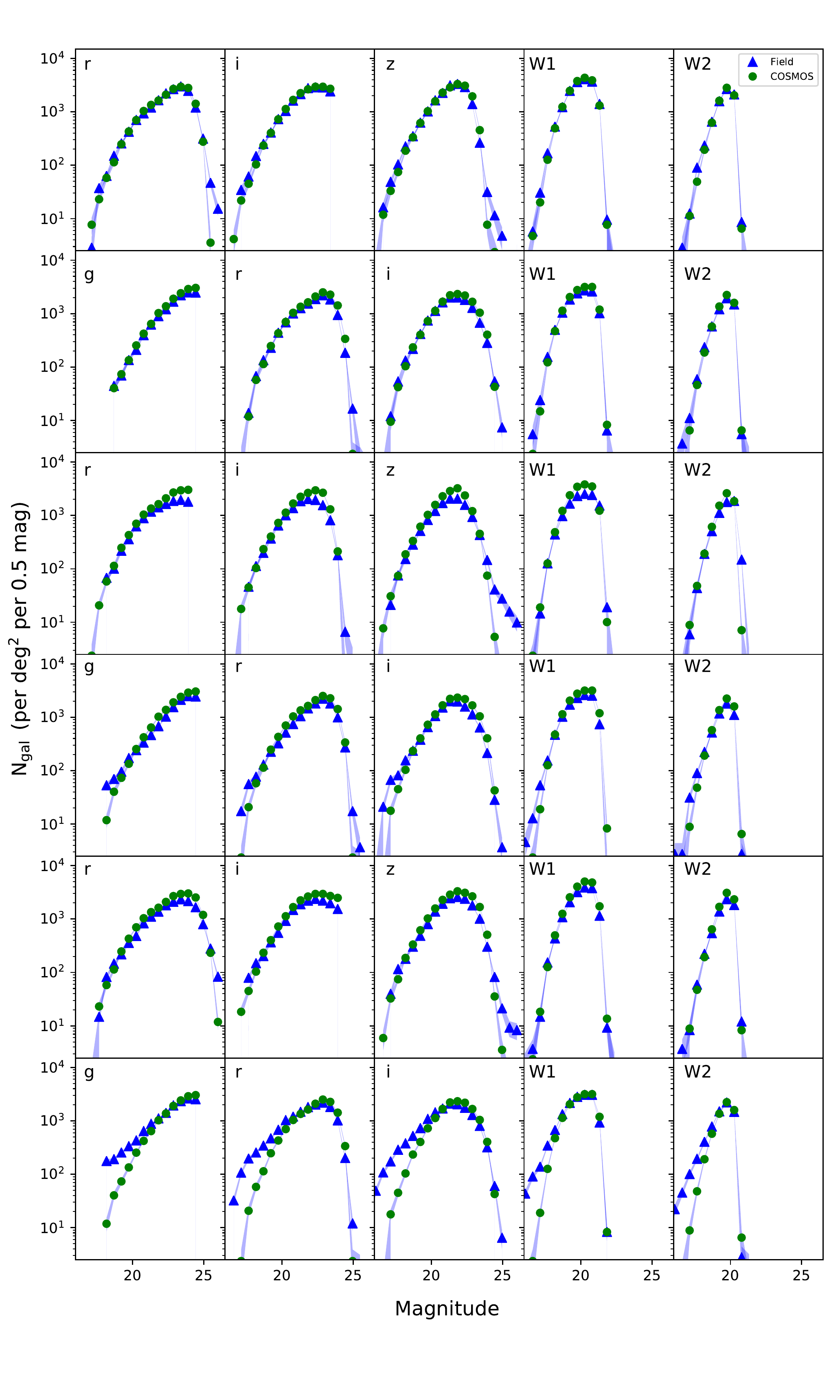}
\caption{\small From top to bottom: galaxy counts for the fields S1 to S6 (blue triangles). COSMOS2015 counts with the same magnitude constraints of each field are also shown (green circles).}
\label{fig:ap_counts}
\end{figure*}

In Figure \ref{fig:ap_counts}, we show the galaxy counts distribution from the S2 to the S6 fields. Each row of plots represents one field and the columns are the different photometric bands. The same magnitude limits adopted for each field were applied to the corresponding COSMOS sample. 

\bsp	
\label{lastpage}
\end{document}